\documentclass[aps,prb,twocolumn,showpacs,preprintnumbers,superscriptaddress]{revtex4}
\usepackage{amssymb}
\usepackage{graphicx}
\usepackage{dcolumn}
\usepackage{bm}
\usepackage{color}

\usepackage{ulem}

\begin{document}

\preprint{PREPRINT (\today)}
\title{Suppression of the Berezinskii-Kosterlitz-Thouless and Quantum Phase Transitions in 2D
Superconductors by Finite Size Effects}
\author{T.~Schneider}
\email{toni.schneider@swissonline.ch}
\affiliation{Physik-Institut der Universit\"{a}t Z\"{u}rich, Winterthurerstrasse 190,
CH-8057 Z\"{u}rich, Switzerland}
\author{S.~Weyeneth}
\affiliation{Physik-Institut der Universit\"{a}t Z\"{u}rich, Winterthurerstrasse 190,
CH-8057 Z\"{u}rich, Switzerland}

\begin{abstract}

We perform a detailed finite-size scaling analysis of the sheet resistance
in Bi-films and the LaAlO$_{3}$/SrTiO$_{3}$ interface in the presence and
absence of a magnetic field applied perpendicular to the system. Our main
aim is to explore the occurrence of Berezinskii-Kosterlitz-Thouless (BKT)
and quantum phase transition behavior in the presence of limited size,
stemming from the finite extent of the homogeneous domains or the magnetic
field. Moreover we explore the implications thereof. Above an extrapolated
BKT transition temperature, modulated by the thickness $d$, gate voltage $%
V_{g}$ or magnetic field $H$, we identify a temperature range where BKT
behavior occurs. Its range is controlled by the relevant limiting
lengths,which are set by the extent of the homogeneous domains or the
magnetic field. The extrapolated BKT transition lines $T_{c}\left(
d,V_{g},H\right) $ uncover compatibility with the occurrence of a quantum
phase transition where $T_{c}\left( d_{c},V_{gc},H_{c}\right) =0$. However,
an essential implication of the respective limiting length is that the
extrapolated phase transition lines $T_{c}\left( d,V_{g},H\right) $ are
unattainable. Consequently, given a finite limiting length, BKT and quantum
phase transitions do not occur. Nevertheless, BKT and quantum critical
behavior is observable, controlled by the extent of the relevant limiting
length. Additional results and implications include: the magnetic field
induced finite size effect generates a flattening out of the sheet
resistance in the $T\rightarrow 0$ limit, while in zero field it exhibits a
characteristic temperature dependence and vanishes at $T=0$ only. The former
prediction is confirmed in both, the Bi-films and the LaAlO$_{3}$/SrTiO$_{3}$
interface, as well as in previous studies. The latter is consistent with the
LaAlO$_{3}$/SrTiO$_{3}$ interface data, while the Bi-films exhibit a
flattening out.
\end{abstract}

\maketitle

%\author{S.~Weyeneth}
%\affiliation{Physik-Institut der Universit\"{a}t Z\"{u}rich, Winterthurerstrasse 190,
%CH-8057 Z\"{u}rich, Switzerland}

%*******************************************

%now the abstract***************************

%*******************************************
%~\\

%
%\narrowtext
%

%*******************************************

\section{Introduction}

Over the last two decades, electrical transport measurements of thin films
near the onset of superconductivity have been studied extensively.\cite%
{markovic,goldrev,gant,goldrev2} Crucial observations include: the sheet
resistance in zero magnetic field remains nearly temperature independent at
the lowest attained temperature \cite{jaeger,goldbi} and remains ohmic below
the expected normal state to superconductor transition temperature $T_{c}$;%
\cite{reyren,caviglia,tsintf} a magnetic field applied perpendicular to the
film generates a flattening out of the sheet resistance in the $T\rightarrow
0$ limit;\cite{ephron,mason,quin,wei} the occurrence of a smeared
Nelson-Kosterlitz jump\cite{nelson} in the superfluid density in the absence%
\cite{turneaure,bert} and presence of a magnetic field.\cite{misra}
Interpretations of the saturation of the sheet resistance in the $%
T\rightarrow 0$ limit include the formation of a metallic phase,\cite%
{ephron,mason,quin,dalido} the occurrence of quantum tunneling of vortices,%
\cite{goldbi,mason} and the failure to cool the electrons.\cite{parendo}

On the other hand, more than three decades ago, Beasley, Mooij, and Orlando\cite{beasley}
suggested that the Berezinskii-Kosterlitz-Thouless\cite{berez,kost} (BKT) transition
may be observable in sufficiently large and
thin superconducting systems. They showed whenever the effective magnetic
penetration depth $\lambda _{2D}=\lambda ^{2}/d$ exceeds the sample size $\left[ W_{s},L_{s}\right] $,
where $\lambda $ is the magnetic penetration
depth, $d$ the thickness, $W_{s}$ the width and $L_{s}$ the length of the
system, the vortices interact logarithmical over the entire sample, a
necessary condition for a BKT transition to occur. Indeed, as shown by
Pearl,\cite{pearl} vortex pairs in thin superconducting systems (charged
superfluid) have a logarithmic interaction energy out to the characteristic
length $\lambda _{2D}=\lambda ^{2}/d$, beyond which the interaction energy
falls off as $1/r$. Accordingly, as $\lambda _{2D}$ increases the
diamagnetism of the superconductor becomes less important and the vortices
in a thin superconducting film become progressively like those in $^{4}$He
films. Invoking the Nelson-Kosterlitz relation\cite{nelson} in the form
$\lambda _{2D}=\lambda ^{2}\left( T_{c}\right) /d=\Phi _{0}^{2}/(32\pi^{2}k_{B}T_{c})$,
it is readily seen that for sufficiently low $T_{c}$'s,
the condition $\lambda _{2D}>\left[ W_{s},L_{s}\right] $ is in practice
accomplishable. Indeed, $T_{c}=1$K yields $\lambda _{2D}\simeq 0.98$ cm.
Additional limiting lengths include the magnetic length $L_{H}\propto \left(
\Phi _{0}/H\right) ^{1/2}$ associated with fields applied perpendicular to
the film and in the case of ac measurements $L_{f}\propto f^{-1/2}$ where $f$
denotes the frequency. Concentrating on dc measurements of the sheet
resistance one expects that the dimension of the homogeneous domains $L_{h}$
sets in zero magnetic field the smallest size so that $L=L_{h}=\min \left[
W_{s},L_{s},\lambda _{2D},L_{h}\right] $. As the magnetic field increases
this applies as long as $L<L_{H}$, while for $L>L_{H}$ the magnetic field
sets the limiting length. It controls the density of free vortices $n_{_{F}}$
which determines the sheet resistance ($R\propto n_{_{F}}$) as well as the
correlation length ($\xi \propto n_{_{F}}^{-1/2}$) at and above $T_{c}$.\cite
{tstool,andersson} Accordingly, the correlation length cannot grow beyond $L$.
In this context it is important to recognize that the finite size
scaling approach adopted here is compatible with the Harris criterion,\cite{harris,aharony}
stating that short-range correlated and uncorrelated
disorder is irrelevant at the BKT critical point, contrary to approaches
where the smearing of BKT criticality is attributed to a Gaussian-like
distribution of the bare superfluid-stiffness around a given mean value.\cite{benefatto}
In this context it should be recognized that irrelevance of this
disorder applies  to the universal properties, while the nonuniversal
parameters, including $T_{c}$ and the vortex core radius, may change.
The finite size effects stemming from the limited extent of the
homogeneous domains or the applied magnetic field have a profound influence
on the observation of the BKT behavior and have been studied intensely in
recent years.\cite{tsintf,tstool,andersson,tswbol,tswbi} On the other hand,
over the years, consistency with BKT behavior has been reported in thin
films,\cite{tswbol,tswbi,hebard,repaci,herbert,crane,weiwei,misra} and in
systems exhibiting interfacial superconductivity.\cite{reyren,caviglia,tsintf,tstool}

Here we extend previous work\cite{tsintf,tstool,tswbol,tswbi} and analyze
the sheet resistance data of Bi-films\cite{goldbi} and the LaAlO$_{3}$/SrTiO$_{3}$
interface\cite{caviglia,tsintf} using the finite size scaling formulas
appropriate for the BKT transition, which include multiplicative corrections
when present.\cite{tstool,andersson} These systems have been chosen because
the data comprise the low temperature limit, namely $T<<T_{c}$ where $T_{c}$
is the extrapolated BKT transition temperature attained in the limit of an
infinite limiting length $L$.

The paper is organized as follows. In Sec. II we sketch the finite size
scaling behavior of the sheet resistance adapted to the BKT critical point
and present the correspondent analysis of the thickness tuned Bi-films and
the gate voltage tuned LaAlO$_{3}$/SrTiO$_{3}$ interface, in the presence
and absence of a magnetic field, applied perpendicular to the film or
interface. We observe remarkable consistency with the finite size scaling
predictions. In the presence and absence of a magnetic field we identify a
temperature range above the extrapolated $T_{c}$ where BKT behavior occurs.
This temperature range is controlled by the relevant limiting length. In
zero magnetic field it is the extent of the homogeneous domains. It turns
out to decrease with the thickness $d$ or gate voltage $V_{g}$ tuned
reduction of $T_{c}\left( d,V_{g}\right) $. The survival of BKT behavior in
applied magnetic fields implies a smeared sudden drop in the superfluid
stiffness at $T_{c}\left( H\right) $, where it adopts the universal value
given by the Nelson-Kosterlitz relation.\cite{nelson} Recently, this
behavior has been observed in MoGe and InO$_{x}$ thin films by means of low
frequency measurements of the ac conductivity.\cite{misra} Analogously,
provided there is a temperature range above $T_{c}\left( d,V_{g}\right)$
where BKT behavior is present, the smeared jump should also occur in zero
field, as observed in various films.\cite{turneaure,bert} An essential
implication of the respective limiting length is that the extrapolated phase
transition lines $T_{c}\left( d,V_{g},H\right)$ are unattainable. As a
consequence the occurrence of BKT transitions is suppressed and with that
the occurrence of quantum phase transitions in the limit $T_{c}\left(
d,V_{g},H\right) \rightarrow 0$ as well. Nevertheless, in agreement with
previous studies,\cite{tsintf,tswbol,tswbi} the lines $T_{c}\left(
d,V_{g},H\right) $ exhibit the characteristic quantum critical properties.
Additional implications of finite size scaling adapted to the BKT transition
include: the magnetic field induced finite size effect generates a
flattening out of the sheet resistance in the $T\rightarrow 0$ limit, while
in zero field it exhibits a characteristic temperature dependence and
vanishes at $T=0$ only. The former prediction is confirmed in both, the
Bi-films and the LaAlO$_{3}$/SrTiO$_{3}$ interface, as well as in previous
studies.\cite{ephron,mason,quin} The latter is consistent with the LaAlO$_{3}
$/SrTiO$_{3}$ interface data, while the Bi-films exhibit a flattening out.
Finally we explore the limitations of the quantum scaling approach.\cite{sondhi}

\section{Theoretical background and data analysis}

\begin{figure}[tbh]
%\centering
\includegraphics[width=1.0\linewidth]{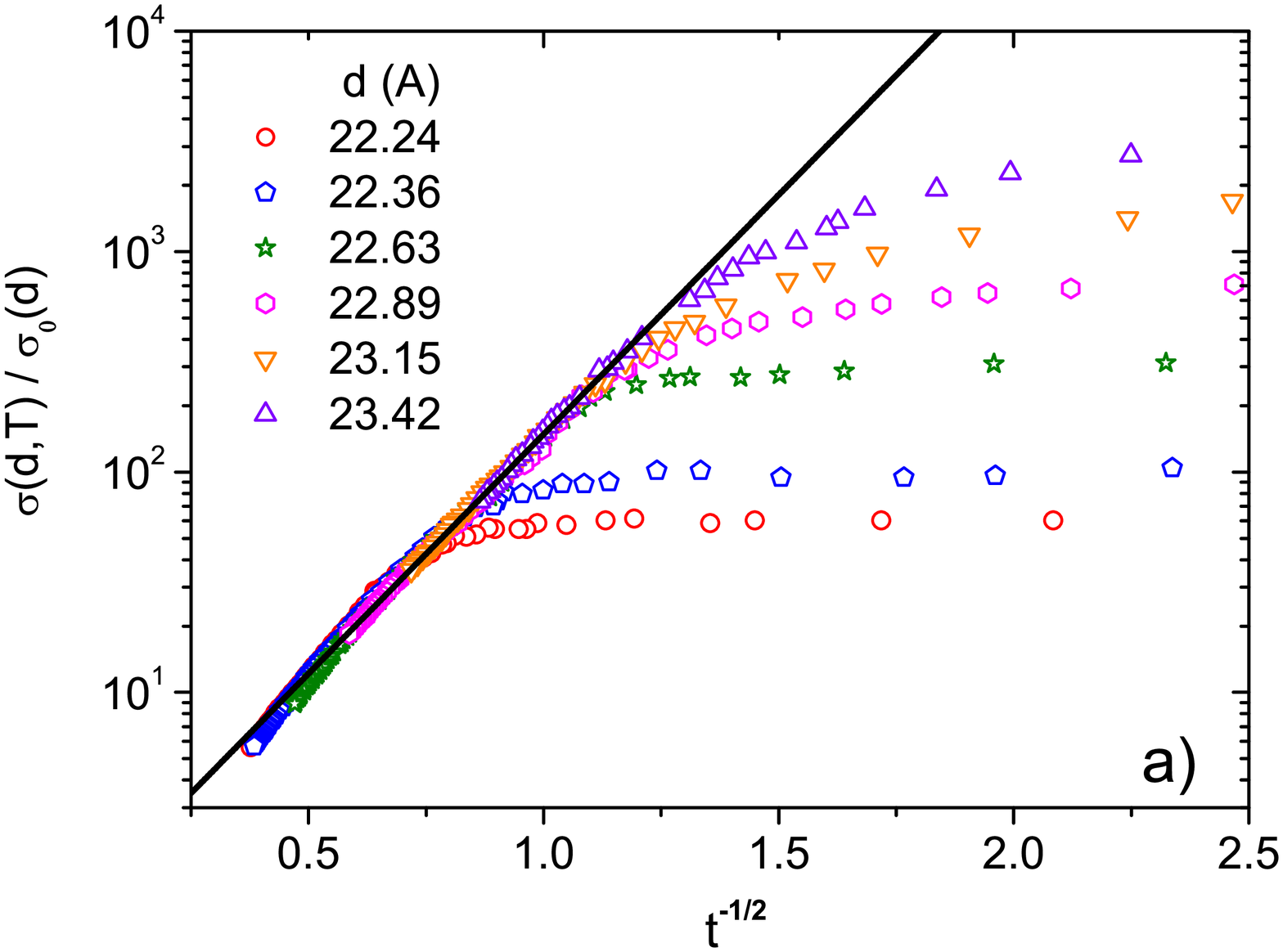} \vspace{-0.5cm} %
\includegraphics[width=1.0\linewidth]{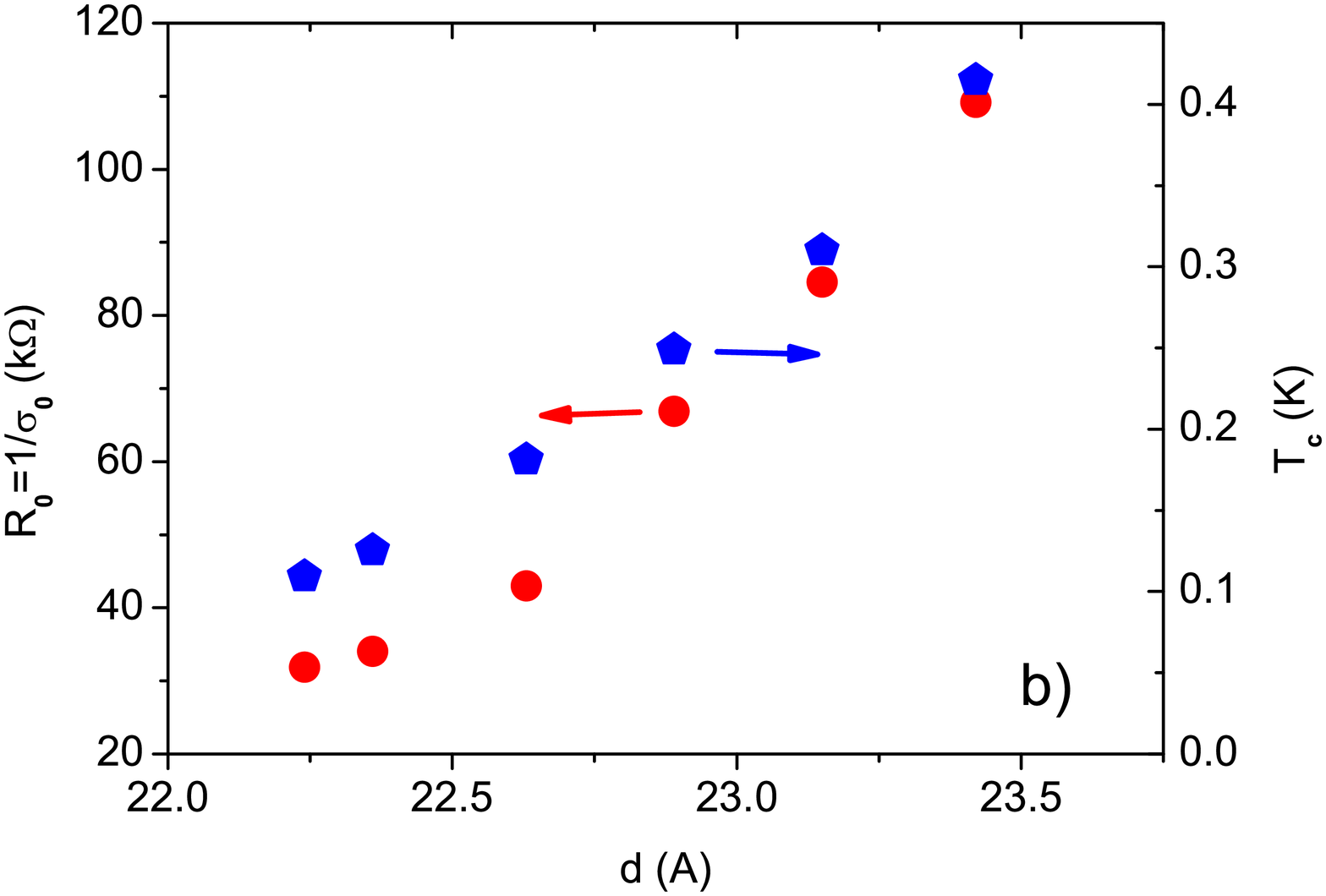} \vspace{-0.5cm}
\caption{(color online)  (a) Normalized sheet conductance $\sigma \left( d,T\right) /\sigma
\left( d\right) $ of Bi films of thickness $d$ vs $%
t^{-1/2}=(T/T_{c}-1)^{-1/2}$ derived from Yen-Hsiang Lin \textit{et al}.\protect\cite{goldbi}
The solid line is the BKT behavior $\sigma \left( d,T\right)
/\sigma _{0}\left( d\right) =$exp$(b_{R}t^{-1/2})$ for a homogenous and
infinite system with $b_{R}=5$. (b) Thickness dependence of the
extrapolated $T_{c}$ and $R_{0}$.}
\label{fig1}
\end{figure}

Since only the motion of free vortices dissipate energy, the sheet
resistance should be proportional to the free vortex density\cite{halperin}

\begin{equation}
R\left( T\right) \propto n_{F}\left( T\right) .  \label{eq1}
\end{equation}%
On the other hand, dynamic scaling predicts the relationship\cite{dsfisher}%
\begin{equation}
R\left( T\right) \propto \xi _{+}^{-z}\left( T\right) ,
\label{eq2}
\end{equation}%
between the sheet resistance above $T_{c}$ and the corresponding correlation
length\cite{ambegaokar}
\begin{equation}
\xi _{+}\left( T\right) =\xi _{0}\exp \left( \frac{2\pi }{bt^{1/2}}\right)
,t=T/T_{c}-1.
\label{eq3}
\end{equation}%
$z$ is the dynamic critical exponent, the amplitude $\xi _{0}$ is related to
the vortex core radius and $b$ is a nonuniversal parameter related to the
vortex core energy.\cite{tsintf,steele} However, approaching $T_{c}$ from
above, the aforementioned limiting lengths imply that the correlation length
$\xi _{+}\left( T\right) $ cannot grow beyond $L=L_{h}=\min \left[
W_{s},L_{s},\lambda _{2D},L_{h}\right] $. According to this a finite size
effect becomes visible around $T^{\ast }>T_{c}$ where%
\begin{equation}
\xi _{+}\left( T^{\ast }\right) \simeq L.
\label{eq4}
\end{equation}%
It leads to a characteristic size dependence of the sheet resistance\cite%
{tsintf,tswbol,tswbi,tstool,andersson} Indeed, Eqs. (\ref{eq2}) and (\ref%
{eq4}) imply that for $z=2$ at $T^{\ast }>T_{c}$ the sheet resistance adopts
the size dependence
\begin{equation}
\frac{\sigma \left( T^{\ast }\right) }{\sigma _{0}}=\frac{R_{0}}{R\left(
T^{\ast }\right) }=\left( \frac{L}{\xi _{0+}}\right) ^{2}
\label{eq5}
\end{equation}
To illustrate the experimental situation we consider next the sheet
resistance data of Yen-Hsiang Lin \textit{et al}.\cite{goldbi} for Bi films
of various thickness and the heat conductance data of Agnolet \textit{et al}.\cite{agnolet}
for a $23.42$ {\AA} thick $^{4}$He film. Both, the sheet
resistance in thin superconducting films and the heat resistance in $^{4}$He
film are supposed to be proportional to the to the free vortex density $n_{F}
$ so that according to Eq. (\ref{eq2}) the respective conductance scales of a
homogeneous film with infinite extent scales for $z=2$ as
\begin{equation}
\frac{\sigma \left( T\right) }{\sigma _{0}}=\frac{R_{0}}{R\left( T\right) }%
=\exp \left( b_{R}t^{-1/2}\right) ,  \label{eq6}
\end{equation}%
where%
\begin{equation}
b_{R}=4\pi /b.  \label{eq7}
\end{equation}%
Supposing that the BKT regime is attainable, $b_{R}$ is nearly independent
of film thickness, $R_{0}$ and $T_{c}$ adopt the appropriate values, the
data plotted as $\sigma \left( T\right) /\sigma _{0}$ vs $t^{-1/2}$
should then fall on the single curve $\exp \left( b_{R}t^{-1/2}\right) $. In
Fig. \ref{fig1}a we depicted this plot for the Bi-films. As $t^{-1/2}$
increases and with that $T_{c}$ is approached the data no longer collapse,
but run away and flatten out at $\sigma \left( T\right) /\sigma _{0}$ values
which increase with film thickness $d$. This behavior points to a finite
size effect where the correlation length $\xi _{+}\left( T\right) $ cannot
grow beyond the limiting length $L$ so that Eq. (\ref{eq5}) applies. As a
result the flattening out is controlled by the ratio $L$ $/\xi _{0+}$ which
increases with film thickness and $T_{c}$ . In Fig. \ref{fig1}b we plotted
the thickness dependence of $R_{0}$ and of the extrapolated BKT transition
line $T_{c}\left( d\right) $. Apparently the decrease of $T_{c}$ with
reduced film thickness points to a quantum phase transition at a critical
thickness dc where $T_{c}\left( d_{c}\right) =0$.. Because the extrapolated
BKT transition temperatures are not attainable due to the limiting length $L,
$ it follows that these transitions, as well as the possible quantum phase
transition at $T_{c}\left( d_{c}\right) =0$ are suppressed. Nevertheless,
slightly above $T_{c}$ , where the data tend to collapse on the BKT line,
BKT fluctuations are present. This collapse attests the consistency with the
universal and characteristic form of the BKT correlation length (Eq. (\ref%
{eq6})), while the nonuniversal parameters $T_{c}$ and $R_{0}$ depend on the
film thickness $d$ (see Fig. \ref{fig1}b). The reduction of $T_{c}$ and $%
R_{0}$ is attributable to disorder and quantum fluctuations. In particular,
the strength of disorder is expected to increase with reduced film thickness
$d$. To quantify this expectation we consider
\begin{equation}
k_{F}l=\left( h/e^{2}\right) /R_{n},  \label{eq7b}
\end{equation}%
where $k_{F}$ denotes the Fermi wavenumber, $l$ the electron mean free path,
and $R_{n}$ the normal state sheet resistance. As disorder increases the
mean free path $l$ diminishes, $k_{F}l$ decreases and the strength of disorder increases. In the Bi-films
considered here $k_{F}l$ varies from $3.8$ for $d=22.2$ {\AA} to $17.4$ for $%
d=23.42$ {\AA}. Accordingly, the strength of the disorder increases
substantially with reduced film thickness or $T_{c}$. Nevertheless, it does
not affect the universal BKT properties but renormalizes the nonuniversal
parameters.

\begin{figure}[tbh]
%\centering
\includegraphics[width=1.0\linewidth]{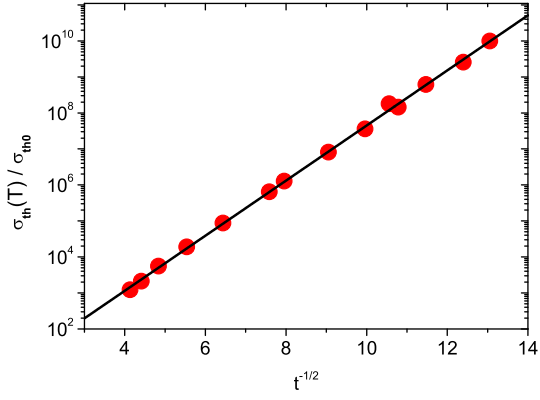} \vspace{-0.5cm}
\caption{(color online) Thermal conduction $\sigma _{th}\left( T\right) /\sigma _{th0}$ of a
$23.42$ \AA\ thick $^{4}He$ film vs $t^{-1/2}$ with $T_{c}=1.2794$ K taken
from Agnolet \textit{et al}.\protect\cite{agnolet} The solid line is the BKT
behavior $\sigma _{th}/\sigma _{th0}=$exp$(b_{R}t^{-1/2})$ with $b_{R}=1.762$
and $\sigma _{th0}=$exp$(-24.13954)=3.283\cdot 10^{-11}$ W/K.}
\label{fig2}
\end{figure}

To classify the relevance of the finite size effect in the Bi-films we show
in Fig. \ref{fig2} the corresponding scaling plot of the thermal conductance of a $%
^{4}$He film. Although the data attain the transition temperature rather
closely there is now sign of a flattening out up to $t^{-1/2}\simeq 13$,
while in the Bi-films it sets in around $0.4\lesssim t^{-1/2}\lesssim 0.75$
(Fig. \ref{fig1}a), depending on the film thickness. Taking this dramatic difference
as a generic fact, a finite scaling analysis of the sheet resistance data
appears to be inevitable to uncover BKT behavior.

\begin{figure}[tbh]
%\centering
\includegraphics[width=1.0\linewidth]{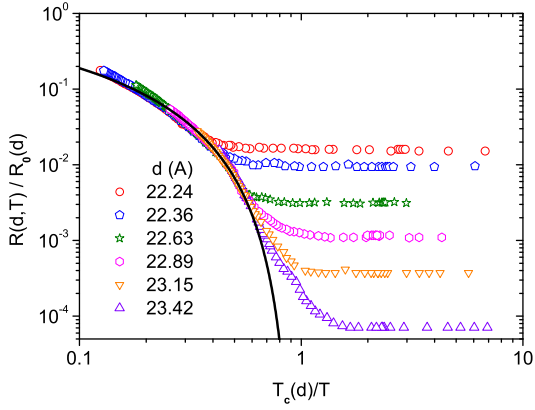} \vspace{-0.5cm}
\caption{(color online)  $R\left( d,T\right) /R_{0}$ vs $T_{c}\left( d\right) /T$ for the
Bi films derived from Yen-Hsiang Lin \textit{et al.\protect\cite{goldbi} }The solid
line is the BKT-behavior $R\left( T\right) /R_{0}=\exp (-b_{R}\left( T/T_{c}-1\right)
^{-1/2})$ with $b_{R}=5$.}
\label{fig3}
\end{figure}

So far we considered finite size effects occurring at and above the transition
temperature $T_{c}$. In Fig. \ref{fig3} we depicted $R\left( d,T\right) /R_{0}$ vs $%
T_{c}\left( d\right) /T$ for the Bi films derived from Yen-Hsiang Lin
\textit{et al}.\cite{goldbi} As $T$ approaches $T_{c}\left( d\right) $ \
the data no longer collapse, but run away from the BKT behavior and flatten
out at $R\left( d,T\right) /R_{0}\left( d\right) $ values which decrease
with film thickness $d$. The flattening out extending above $T_{c}\left(
d\right) /T>1$ points then to a finite size effect below $T_{c}\left(
d\right) $ as well. However, below $T_{c}$ the dynamic scaling relation (\ref%
{eq2}) is no longer applicable because the correlation length is infinite
there owing to the divergence of the susceptibility.\cite{kost}

The BKT theory predicts that below $T_{c}$ all vortices are bound
in pairs by the logarithmic vortex interaction, whereupon the linear sheet
resistance is zero. Instead there is a nonlinear dependence of the voltage
on current since the current can unbind weakly bound pairs.\cite{halperin}
Contrariwise, in a finite sample there will be a population of free vortices
at and below the vortex unbinding transition temperature $T_{c}$.\cite%
{andersson,repaci,herbert} In this temperature regime the linear
relationship (\ref{eq1}) between sheet resistance and free vortex density
still applies, while Eq. (\ref{eq2}), relating the sheet resistance to the
correlation length (Eq. (\ref{eq3})), applies at and above $T_{c}$ only. To
provide a rough estimate of the free vortex density we note that at low
temperatures the energy change resulting from adding a single vortex in a
system of size $L$ is given by $\Delta E=\left( J\left( T\right) /2\right)
\int_{0}^{2\pi }d\Theta \int_{\xi _{0}}^{L}RdR/R^{2}=\pi J\left( T\right)
\ln \left( L/\xi _{0-}\right) $,\cite{jensen} where $\xi _{0}$ is the vortex core radius and

\begin{equation}
J\left( T\right) =\hbar ^{2}\rho _{s}\left( T\right) /2m=d\Phi
_{0}^{2}/\left( 16\pi ^{3}\lambda ^{2}\left( T\right) \right) ,  \label{eq7a}
\end{equation}%
denotes the superfluid stiffness at low temperatures ($T<<T_{c}$). An
estimate for the free vortex density follows then from the probability of
finding a free vortex from the Boltzmann factor\begin{equation}
P\left( T\right) \propto n_{F}\left( T\right) \propto \exp (-\Delta
E/k_{B}T)=\left( \xi _{0}/L\right) ^{\pi J\left( T\right) /k_{B}T}.
\label{eq8}
\end{equation}%
Using Eq. (\ref{eq1}) we obtain,%
\begin{equation}
R\left( T\right) \propto n_{F}\left( T\right) \propto \left( \xi
_{0}/L\right) ^{\pi J\left( T\right) /k_{B}T}:T<<T_{c}.  \label{eq9}
\end{equation}%
Invoking the universal Nelson-Kosterlitz relation\cite{nelson}%
\begin{equation}
k_{B}T_{c}=\frac{\pi }{2}J\left( T_{c}^{-}\right) ,  \label{eq10}
\end{equation}%
the temperature range of validity is then restricted to $T<<T_{c}=\pi
J\left( T_{c}^{-}\right) /2k_{B}$. As it should be, for an infinite system, $%
n_{F}$ is zero for $T\leq T_{c}$. But if the limiting length $L$ is finite,
the free vortex density vanishes at zero temperature only. This implies an
ohmic tail in the IV characteristic below the extrapolated $T_{c}$\cite%
{repaci,herbert,reyren} and impedes a normal state to superconductor
transition at finite temperature in a strict sense. In this context it is
important to recognize that the standard finite size scaling outlined above
neglects the multiplicative logarithmic corrections associated with BKT
critical behavior.\cite{andersson,medved} A recent renormalization group
treatment yields for $z=2$ and free boundary conditions\cite{andersson}%
\begin{equation}
R\left( T\right) \propto \left\{
\begin{array}{c}
\left( \xi _{0}/L\right) ^{\pi J\left( T\right) /k_{B}T}\text{ \ \ \ \ \ \ \
\ \ \ \ \ \ \ \ : }L\gtrsim \xi _{-}\left( T\right) \\
\left( \xi _{0}/L\right) ^{2}/\ln \left( \left( L_{\lim }/\xi _{0}\right)
/b_{0}\right) \text{ : }L\lesssim \xi _{+}\left( T\right)%
\end{array}\right.,
\label{eq11}
\end{equation}%
where%
\begin{equation}
\xi _{-}\left( T\right) =\xi _{0}\exp \left( \frac{1}{b\left\vert
t\right\vert ^{1/2}}\right) ,  \label{eq12}
\end{equation}%
is a diverging length below $T_{c}$.\cite{ambegaokar} With Eq. (\ref{eq3}) it
follows that this thermal length is much smaller than the correlation length
$\xi _{+}\left( T\right) $ for the same $\left\vert t\right\vert $, because
\begin{equation}
\xi _{+}\left( t\right) /\xi _{0}=\left( \xi _{-}\left( \left\vert
t\right\vert \right) /\xi _{0}\right) ^{2\pi }.  \label{eq12a}
\end{equation}%
The parameter $b_{0}$ is fixed by the initial conditions of the
renormalization group equations,\cite{andersson} while the derivation of Eq.
(\ref{eq9}) identifies $\xi _{0}$ as vortex core radius. Furthermore, there
is the upper bound $b_{0}<L/\xi _{0-}$ because $R\left( T\right) >0$.
Taking the multiplicative logarithmic correction into account Eq.( \ref{eq5}) transforms with Eq. (\ref{eq11}) to%
\begin{equation}
\frac{R\left( T_{c}\right) }{R_{0}}=\frac{\sigma _{0}}{\sigma \left(
T_{c}\right) }=\left( \frac{\xi _{0}}{L}\right) ^{2}\frac{1}{\ln \left(
\left( L/\xi _{0}\right) /b_{0}\right) },  \label{eq13}
\end{equation}%
valid at $T\simeq T_{c}$.

\begin{figure}[tbh]
%\centering
\includegraphics[width=1.0\linewidth]{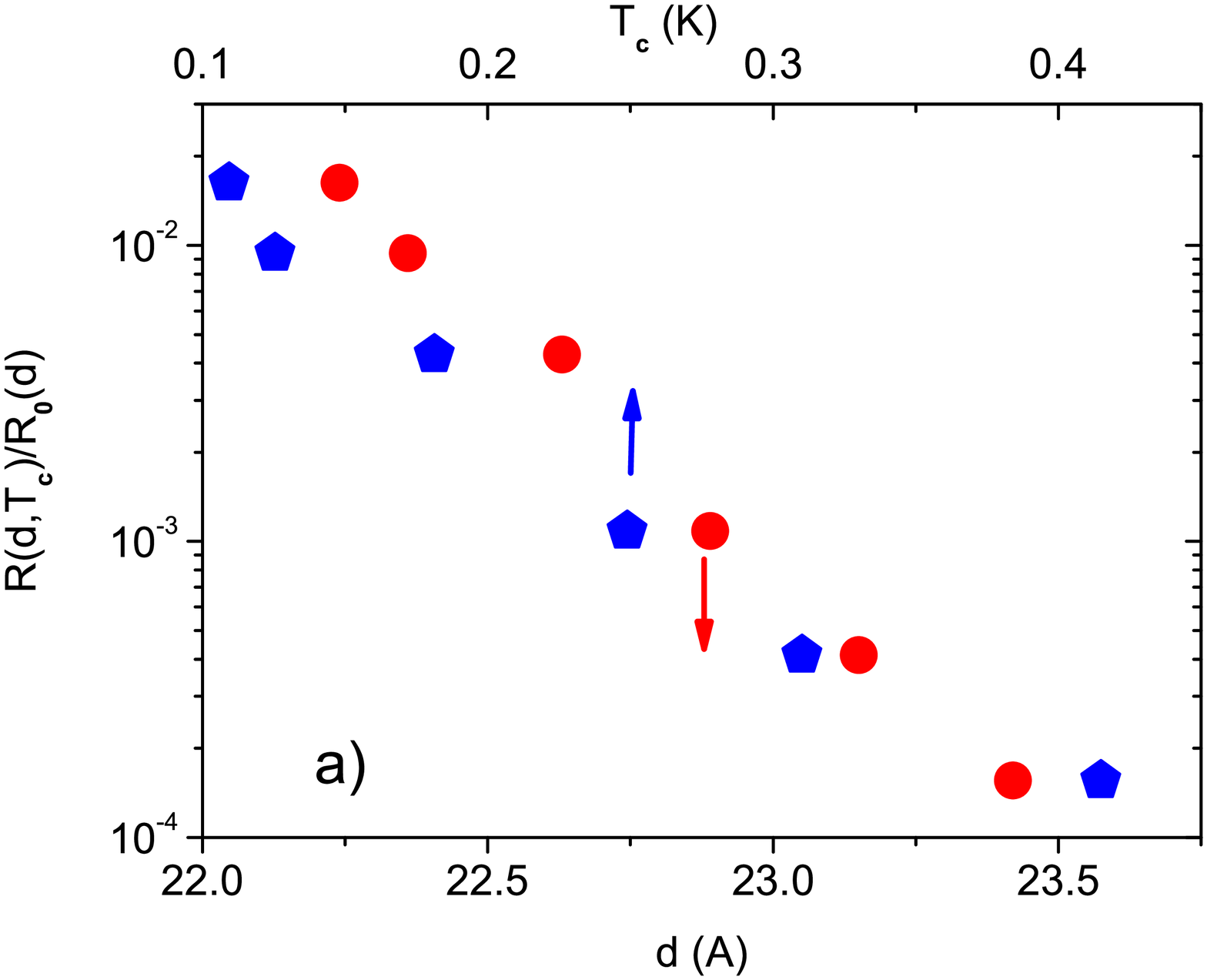} \vspace{-0.5cm} %
\includegraphics[width=1.0\linewidth]{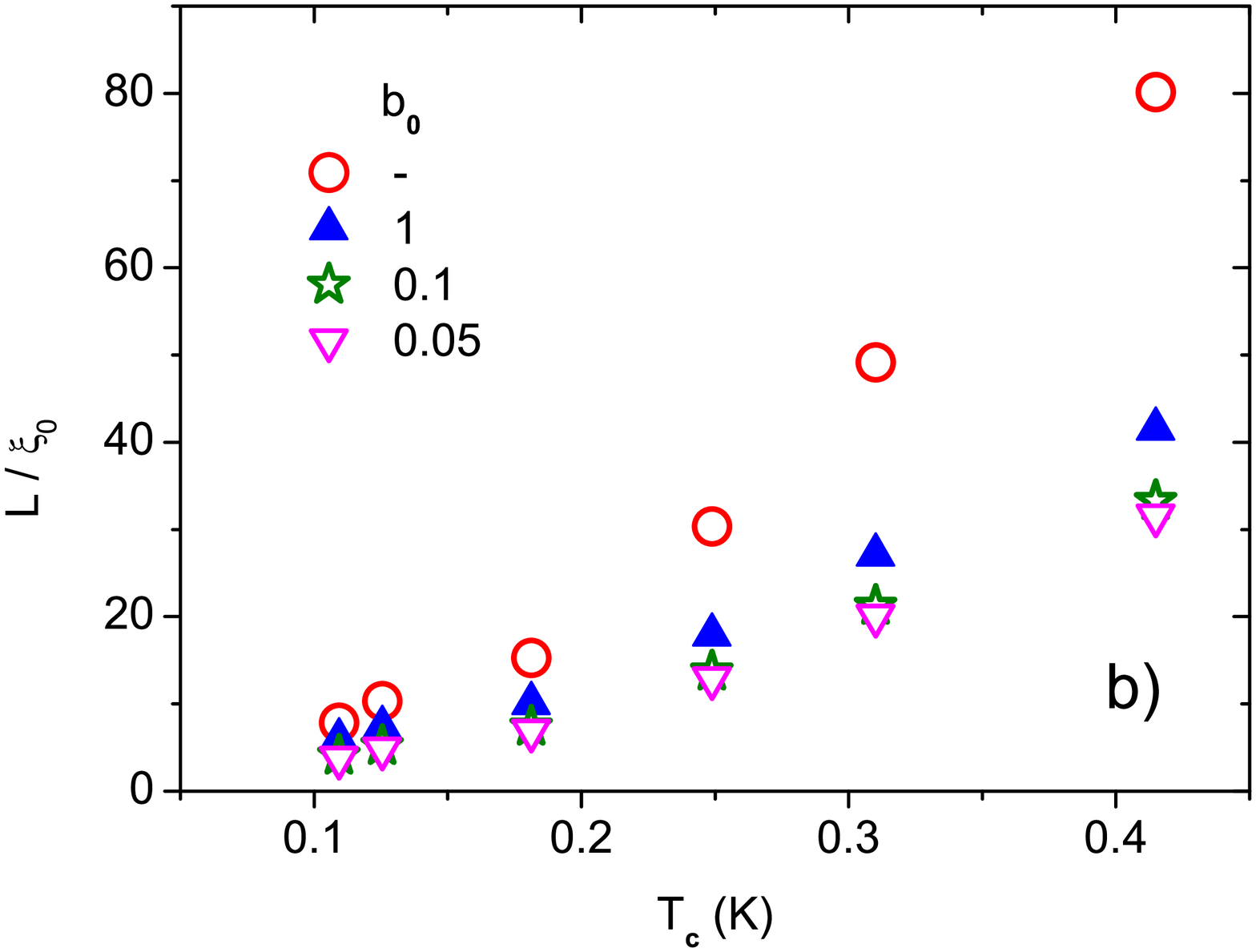} \vspace{-0.5cm}
\caption{(color online)  (a) $R\left( d,T\right) /R_{0}\left( d\right) $ vs $T_{c}$
and $d$ derived from the data shown in Fig. \protect\ref{fig3}; (b) Estimates for the ratio $%
L/\xi _{0}$ between correlation length and vortex core radius without the
multiplicative logarithmic correction term ($\bigcirc $) and with this
correction for different $b_{0}$ values entering Eq. \protect\ref{eq13}.}
\label{fig4}
\end{figure}

Given $R\left( T_{c}\right) /R_{0}$ and $b_{0}$, estimates for $L_{\lim
}/\xi _{0-}$ are then readily obtained. Fig. \ref{fig4}a depicts the $T_{c}$ and $d$
dependence of $R\left( T_{c}\right) /R_{0}$ derived from Fig. \ref{fig3}, and the
resulting $T_{c}$ dependence of $L_{\lim }/\xi _{0-}$ is shown in Fig. \ref{fig4}b
for $b_{0}=0.05, 0.1$ and $1$ in comparison with the neglect of the
multiplicative logarithmic correction. These $b_{0}$ values satisfy the
lower bound $b_{0}<L/\xi _{0}$ resulting from the requirement, $R\left(
d,T_{c}\right) /R_{0}\left( d\right) >0$.  Furthermore, $b_{0}=0.05$ is
comparable to $b_{0}\approx 0.07$, derived from large-scale numerical
simulations.\cite{andersson} Striking features include the substantial
decline of the ratio between limiting length and vortex core radius, $L/\xi
_{0}$, with decreasing $T_{c}$, and the comparably low $L/\xi _{0}<80$
values. Indeed, the run away is controlled by the magnitude of $L/\xi _{0}$.
The $^{4}$He data shown in Fig. \ref{fig2} do not exhibit a sign of flattening out up
to $\sigma _{th}\left( T\right) /\sigma _{th0}$ $=10^{10}$, yielding with Eq. (\ref%
{eq5}) the lower bound $L/\xi _{0}\gtrsim 10^{5}$. According to this, the
run away observed in Fig. \ref{fig1}a and Fig. \ref{fig3} stems from a limiting length $L$ where
the ratio $L/\xi _{0}$ decreases with film thickness. Nevertheless, there is
a temperature range where consistency with BKT behavior is observed, but in
a strict sense a normal state to superconductor BKT transition is
suppressed. As a consequence , there is also no film thickness driven
quantum phase transition where the phase transition line $T_{c}\left(
d\right) $ ends at $T_{c}\left( d_{c}\right) =0$ vanishes at a critical film
thickness $d_{c}$, as could be anticipated from the thickness dependence of
the extrapolated $T_{c}$ shown in Fig. \ref{fig1}b.

\begin{figure}[tbh]
%\centering
\includegraphics[width=1.0\linewidth]{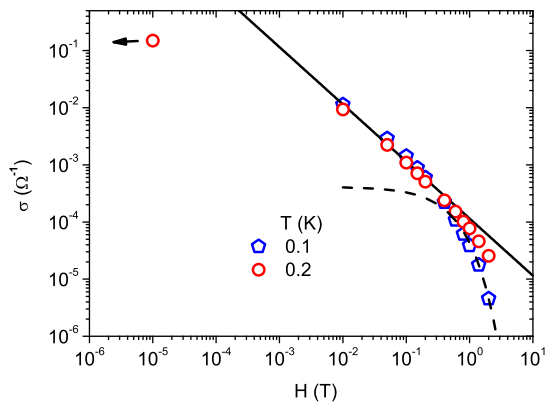} \vspace{-0.5cm}
\caption{(color online) Sheet conductivity $\sigma $ of the $23.42$ {\AA} thick Bi-film vs magnetic
field $H$ at $T=0.1$ K and $0.2$ K derived from Yen-Hsiang Lin
\textit{et al}.\protect\cite{goldbi} The solid line is Eq. \protect\ref{eq15} in the from $%
\sigma =\widetilde{\sigma }/H$ where $\widetilde{\sigma }=1.15\cdot 10^{-4}$
($\Omega ^{-1}$T). The dashed line is Eq. \protect\ref{eq18} with $\overline{%
\sigma }_{0}=4.12\cdot 10^{-4}$ $\Omega ^{-1}$ and $\overline{a}=2.25$ T$%
^{-1}$. The arrow indicates that this data point marks the zero field value
of the sheet conductivity.}
\label{fig5}
\end{figure}

An essential issue left is the elucidation of the limiting length $L_{\min }$.
In principle the magnetic field induced finite size effect offers a direct
estimate. A magnetic field applied perpendicular to the film leads to the
limiting length\cite{tstool}%
\begin{equation}
L_{H}=\left( \frac{\Phi _{0}}{aH}\right) ^{1/2},  \label{eq14}
\end{equation}
where $a\approx 4.8$ fixes the mean distance between vortices. It prevents
the divergence of the correlation length at the extrapolated $T_{c}$. In
analogy to Eq. (\ref{eq5}) the sheet resistance is then expected to scale
as
\begin{equation}
R(H,T_{c})=\frac{1}{\sigma (H,T_{c})}=\frac{f}{L_{H}^{2}}=\frac{afH}{\Phi
_{0}},  \label{eq15}
\end{equation}
for $z=2$ and low fields applied perpendicular to the film.\cite%
{tstool,andersson} In contrast to the zero field scaling form (\ref{eq11}),
this law holds below $T_{c}$ as well and the additive correction to the
leading power law dependence is weak.\cite{andersson} The magnetic field
induced finite sets then the limiting length as long as $L_{H}\propto
H^{-1/2}<L$ whereby $L_{H}$ increases with decreasing field and approaches $L
$. Here a runaway from the scaling behavior (\ref{eq15}) sets in at $H^{\ast
}$ providing for $L$ the estimate%
\begin{equation}
L=\left( \frac{\Phi _{0}}{aH^{\ast }}\right) ^{1/2}.  \label{eq16}
\end{equation}%
In Fig. \ref{fig5} we depicted the magnetic field dependence of the sheet
conductivity of the $23.42$ {\AA} thick Bi film at $T=0.1$ K and $0.2$ K
where the latter is close to the extrapolated $T_{c}$. Even though the data
are rather sparse we observe in a intermediate magnetic field range
consistency with the predicted linear and nearly temperature independent
field dependence of the sheet resistance. No, we focus on the low field
behavior of the conductivity shown in Fig. \ref{fig5}. The run away from the $1/H$
dependence of the sheet conductivity occurs around $H=0.01$ T$\simeq H^{\ast
}$, yielding with Eq. (\ref{eq16}) for the limiting length the estimate%
\begin{equation}
L\simeq 208\text{{\AA}.}  \label{eq17}
\end{equation}%
With $L_{\min }/\xi _{0}\simeq 32$, taken from Fig. \ref{fig4}b, we obtain for the
magnitude of the radius of the vortex core radius
\begin{equation}
\xi _{0}\simeq 6.5\text{ {\AA}.}  \label{eq18}
\end{equation}%
The deviations from the finite size scaling behavior at higher fields are
not unexpected because with increasing magnetic field the BKT regime is
gradually left and he isotherms cross around $H=H_{c}\simeq 0.4$ T,
signaling the occurrence of a magnetic field driven quantum phase
transition. In addition Eq. (\ref{eq15}) captures the leading field
dependence only. In the field range where it applies the plot $\sigma $ vs $%
1/H$ shown in Fig. \ref{fig5} also reveals a nearly temperature independent
coefficient of proportionality $\widetilde{\sigma }$. It implies that the
temperature dependence of the sheet resistance at fixed field flattens out,
as observed in the $23.42$ {\AA} thick Bi film,\textit{\cite{goldbi} } Analogous
behavior was also observed in MoGe films,\cite{mason} and Ta-films.\cite%
{quin} in a field range where the magnetic field induced finite size scaling
approach is no longer applicable. Indeed, in the MoGe films the temperature
independent sheet resistance obeys the empirical form\cite{mason}%
\begin{equation}
\sigma \left( H\right) =\overline{\sigma }_{0}\exp \left( -\overline{a}%
H\right) .  \label{eq18a}
\end{equation}%
In the present case it applies according to Fig. \ref{fig5} at best above the
critical field only. The unusual empirical form was attributed to
dissipative quantum tunneling of vortices from one "insulating" patch to
another.

\begin{figure}[tbh]
%\centering
\includegraphics[width=1.0\linewidth]{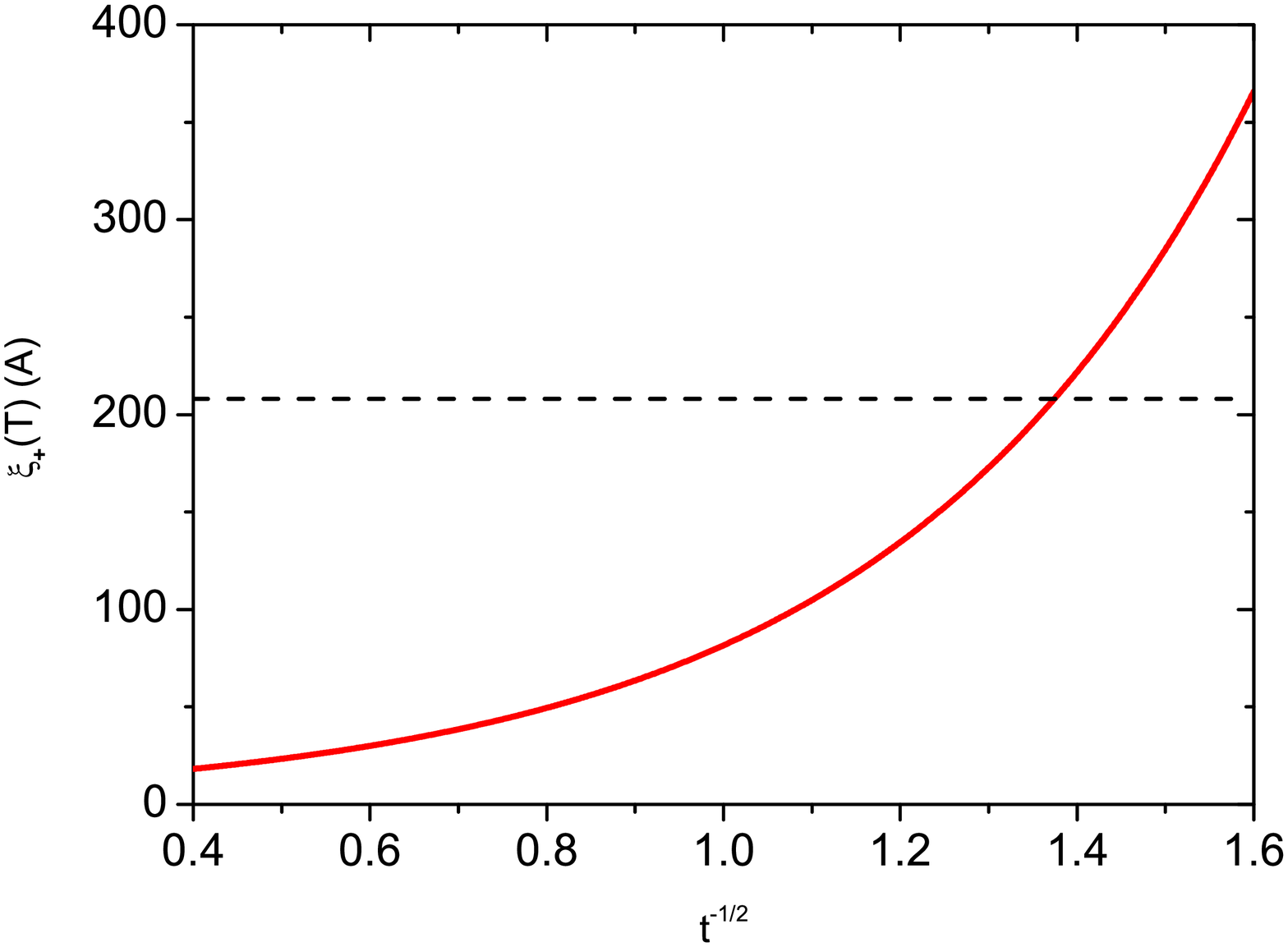} \vspace{-0.5cm}
\caption{(color online) Correlation length
$\xi _{+}=\xi _{0}\exp \left( 2\pi \left( bt\right) ^{-1/2}\right) $ vs $%
t^{-1/2}$ of the $23.42$ {\AA} thick Bi-film with $\xi _{0}=6.5$ {\AA} and $2\pi
/b=b_{R}/2=2.5$. The dashed line marks $L=208$ {\AA}. The crossing point at $%
t^{-1/2}\simeq 1.38$ corresponds to $T_{c}/T\simeq 0.66$.}
\label{fig6}
\end{figure}

As the estimates for $L_{\min }$ and $\xi _{0}$ stem from rather sparse data
a reliability check is inevitable. For this purpose we consider the
temperature dependence of the correlation length $\xi _{+}$ (Eq. (\ref{eq3}%
)) of the $23.42$ {\AA} thick Bi film in terms of $\xi _{+}\left( T\right) $ vs $t^{-1/2}$ with $\xi _{0}=6.5$ {\AA} shown in Fig. \ref{fig6}. As $\xi _{+}$ growth with
increasing $t^{-1/2}$ it approaches the limiting length $L=208$ {\AA} at $%
t^{-1/2}\simeq 1.38$, the range where in this film the run away from BKT
behavior occurs (see Fig. \ref{fig1}a). Accordingly, we established for the $23.42$ {\AA}
thick Bi-film reasonable consistency between the estimates for the vortex
core radius $\xi _{0}$ and the limiting length $L$, derived from the
magnetic field induced finite size effect, and the observed zero field
behavior of the sheet resistance. Unfortunately, this estimation of $\xi
_{0}$ and $L$ is restricted to this film because the magnetic field
dependence of the sheet resistance appears to be missing for the other
films. In any case, the rather small limiting length $L=208$ {\AA} points to
an inhomogeneous film, with homogeneous patches of dimension $L=L_{h}$.

\begin{figure}[tbh]
%\centering
\includegraphics[width=1.0\linewidth]{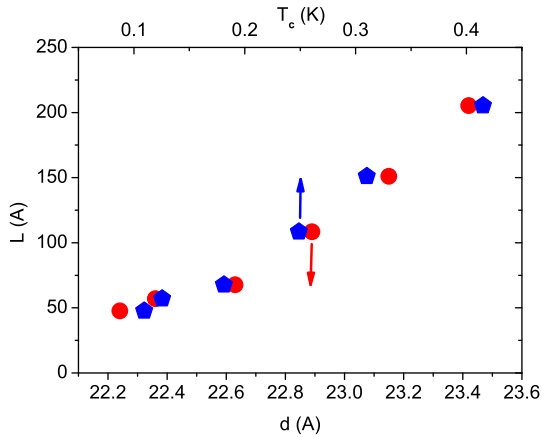} \vspace{-0.5cm}
\caption{(color online) $T_{c}$ and film thickness dependence of the limiting
length $L$ of the Bi-films derived from the $L/\xi _{0}$ estimates shown in
Fig. \protect\ref{fig4}b for $b_{0}=0.05$ and $\xi _{0}=gT_{c}^{-1/2}$ with $g=4.19$ {\AA}K$%
^{1/2}$.}
\label{fig7}
\end{figure}

In this context it should be kept in mind that there is the Harris criterion,%
\cite{harris,aharony} stating that short-range correlated and uncorrelated
disorder is irrelevant at the unperturbed critical point, provided that $\nu >2/D$,
where $D$ is the dimensionality of the system and $\nu $
the critical exponent of the finite-temperature correlation length. With $D=2
$ and $\nu =\infty $, appropriate for the BKT transition,\cite{kost}
this disorder should be irrelevant. Given the irrelevance of disorder, the
reduction of the ratio $L/\xi _{0}$ with reduced film thickness or
transition temperature (see Fig.\ref{fig4}b) is then attributable to: (i) increasing
vortex core radius $\xi _{0}$ with reduced $T_{c}$ combined with a thickness
independent $L$; (ii) a limiting length $L$ which decreases with film
thickness combined with a $T_{c}$ independent $\xi _{0}$; (iii) a thickness
dependence of both, $L$ and $\xi _{0}$, such that the ratio $L/\xi _{0}$
decreases with reduced transition temperature. Because the vortex core
radius is known to increase with reduced $T_{c}$ as $\xi _{0}\propto
T_{c}^{-1/z}$ with $z=2$,\cite{finotello,williams} we are left with option
(i) and (iii). In order to discriminate between these options we estimate $%
\xi _{0}\left( T_{c}\right) $ from the respective data for the $23.42$ {\AA}
thick Bi film, namely $\xi _{0}=6.5$ {\AA} and $T_{c}=0.41$ K, yielding $\xi
_{0}=gT_{c}^{-1/2}$ with $g=4.19$ {\AA}K$^{1/2}$. The rough estimates for the
thickness and $T_{c}$ dependence of $L$ shown in Fig.\ref{fig7} are then readily
obtained from the $L/\xi _{0}$ values depicted in Fig. \ref{fig4}b. In spite of the
small total thickness increment of $1.18$ {\AA} there is a strong
thickness dependence of $L$, ranging from $50$ {\AA} to $200$ {\AA}. Direct
experimental evidence for superconducting patches with an extent of $100$ {\AA}
embedded in an insulating background stems from scanning tunneling
spectroscopy investigations on TiNi\cite{sacepe} and InO$_{x}$\cite{sacepe2}
films. However, it should be kept in mind that transport measurements are
sensitive to the phase and tunneling experiments to the magnitude of the
order parameter. Furthermore, scanning SQUID measurements at the interface
LaAlO$_{3}$/SrT iO$_{3}$ uncovered superconducting regions occupying only a
small fraction of the areas measured. In addition there are magnetic regions
with patches of ferromagnetic regions coexisting with a higher density of
much smaller scale domains of fluctuating local magnetic moments.\cite{bert2}

\begin{figure}[tbh]
%\centering
\includegraphics[width=1.0\linewidth]{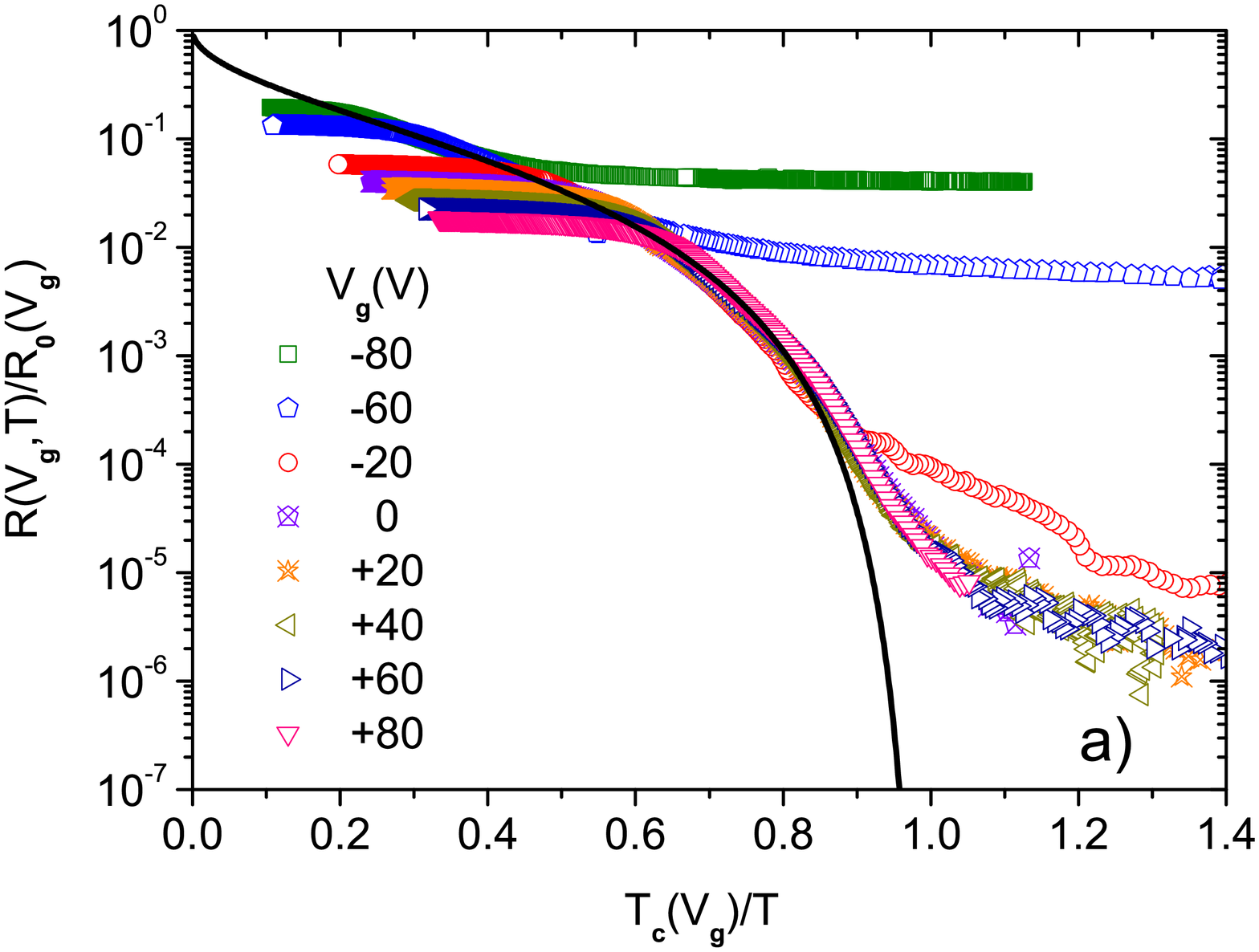} \vspace{-0.5cm}
\includegraphics[width=1.0\linewidth]{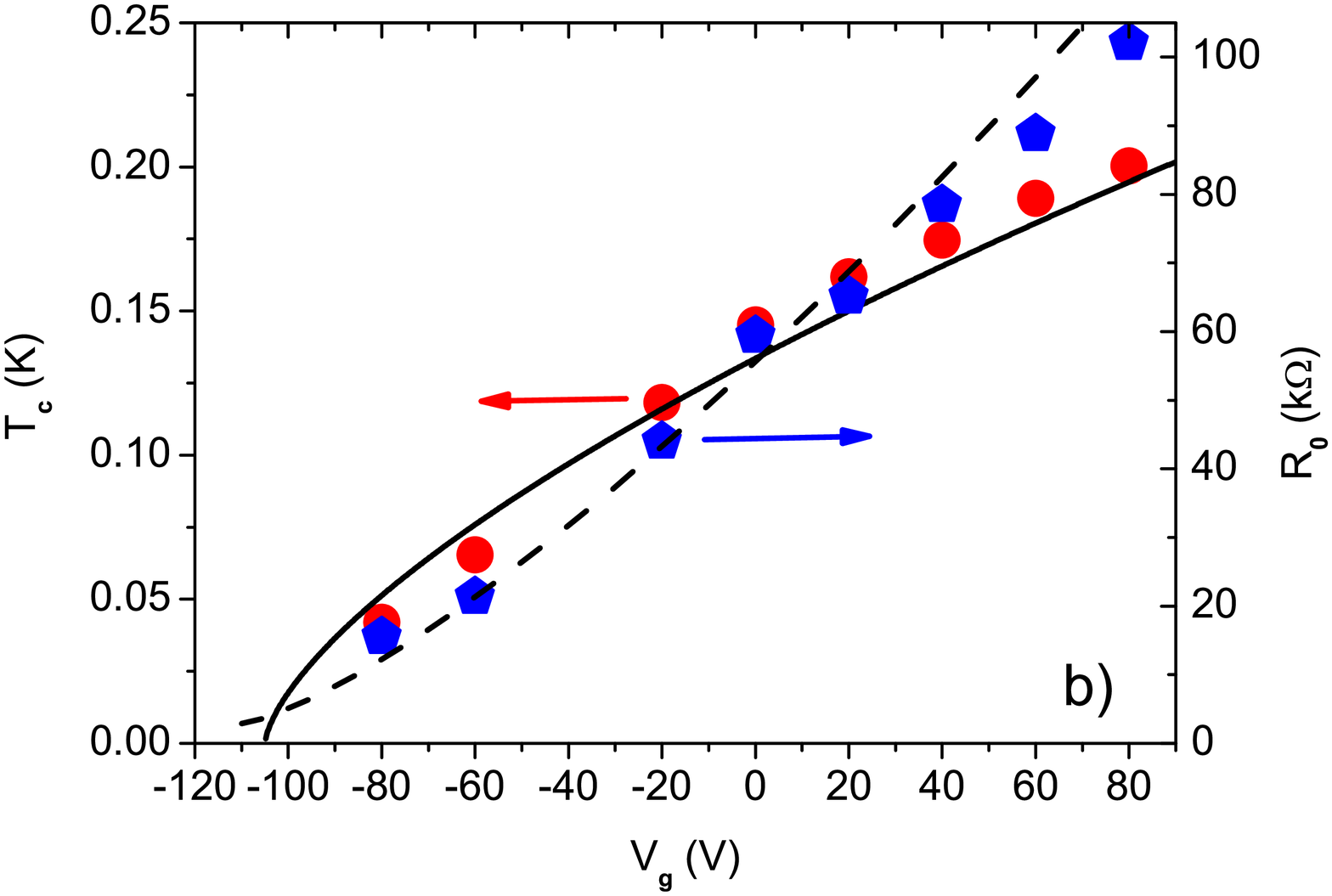} \vspace{-0.5cm}
\caption{(color online) (a) Normalized sheet resistance $R\left( V_{g},T\right)
/R_{0}\left( V_{g}\right) $ vs $T_{c}\left( V_{g}\right) /T$ of the LaAlO$%
_{3}$/SrTiO$_{3}$ interface at various gate voltages derived from Caviglia
\textit{et al.\protect\cite{caviglia} }The solid line marks the BKT behavior $%
R\left( V_{g},T\right) /R_{0}\left( d\right) =\exp \left( -b_{R}\left(
T/T_{c}-1\right) ^{-1/2}\right) $ for a homogenous and infinite system with
with $b_{R}=3.43$. (b) Gate voltage dependence of the extrapolated transition
line $T_{c}\left( V_{g}\right) $ and $R_{0}\left( V_{g}\right) $. The solid
and dashed lines indicate the approach of $T_{c}$ and $R_{0}$ to the
extrapolated quantum phase transition.}
\label{fig8}
\end{figure}

To explore the finite size scenario further we turn to the interface between
LaAlO$_{3}$ and SrTiO$_{3}$, two excellent band insulators. It was shown
that the electric-field effect can be used to map the phase diagram of this
interface system revealing, depending on the gate voltage, a smeared BKT
transition and evidence for quantum critical behavior.\cite{caviglia,tsintf}
Here we revisit the analysis of the temperature and gate voltage dependence
of the sheet resistance data by invoking the approach outlined above. In
Fig. \ref{fig8}a we depicted $R\left( V_{g},T\right) /R_{0}$ vs $T_{c}\left(
V_{g}\right) /T$ and in Fig. \ref{fig8}b the gate voltage dependence of the
extrapolated transition temperature $T_{c}$ and amplitude $R_{0}$. As $%
T_{c}\left( V_{g}\right) /T$ increases Fig. \ref{fig8}a uncovers a flow to and away
from the BKT behavior. As $T_{c}\left( V_{g}\right) /T$
decreases for fixed $T_{c}$ the BKT regime is left, while the rounding of
the transition leads with increasing $T_{c}\left( V_{g}\right) /T$ to a
flow away from criticality. Nevertheless, in an intermediate $T_{c}\left(
V_{g}\right) /T$ regime the data tend to collapse on the characteristic
BKT line. Thus, in analogy to the Bi-films, the collapse attests again
consistency with the universal and characteristic form of the BKT
correlation length (Eq. (\ref{eq6})), while the nonuniversal parameters $%
T_{c}$ and $R_{0}$ depend in the present case on the gate voltage (see Fig. %
\ref{fig8}b). Their reduction points to the occurrence of a gate voltage
tuned quantum phase transition around $Vg\simeq -100$ V where the
extrapolated transition temperature vanishes. Using Eq. (\ref{eq7b}) we find
that $k_{F}l$ varies from $8.5$ at $V_{g}=80$ V to $13.7$ for $V_{g}=+80$ V.
Accordingly, disorder is present, its strength is comparable to that in the
Bi-films but increases only slightly by approaching the extrapolated quantum
phase transition. In any case, it does not affect the universal BKT
properties but renormalizes the nonuniversal parameters.

\begin{figure}[tbh]
%\centering
\includegraphics[width=1.0\linewidth]{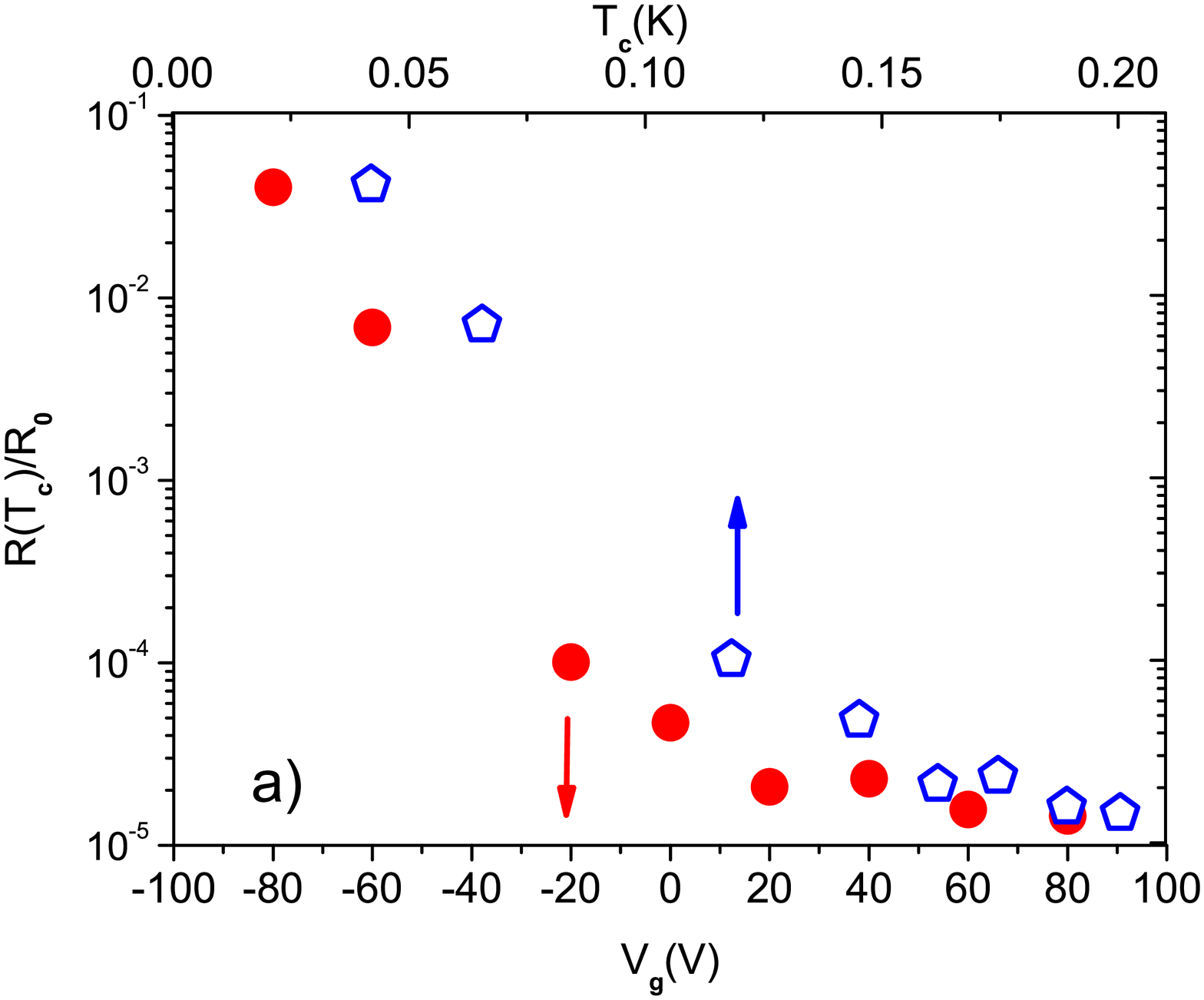} \vspace{-0.5cm}
\includegraphics[width=1.0\linewidth]{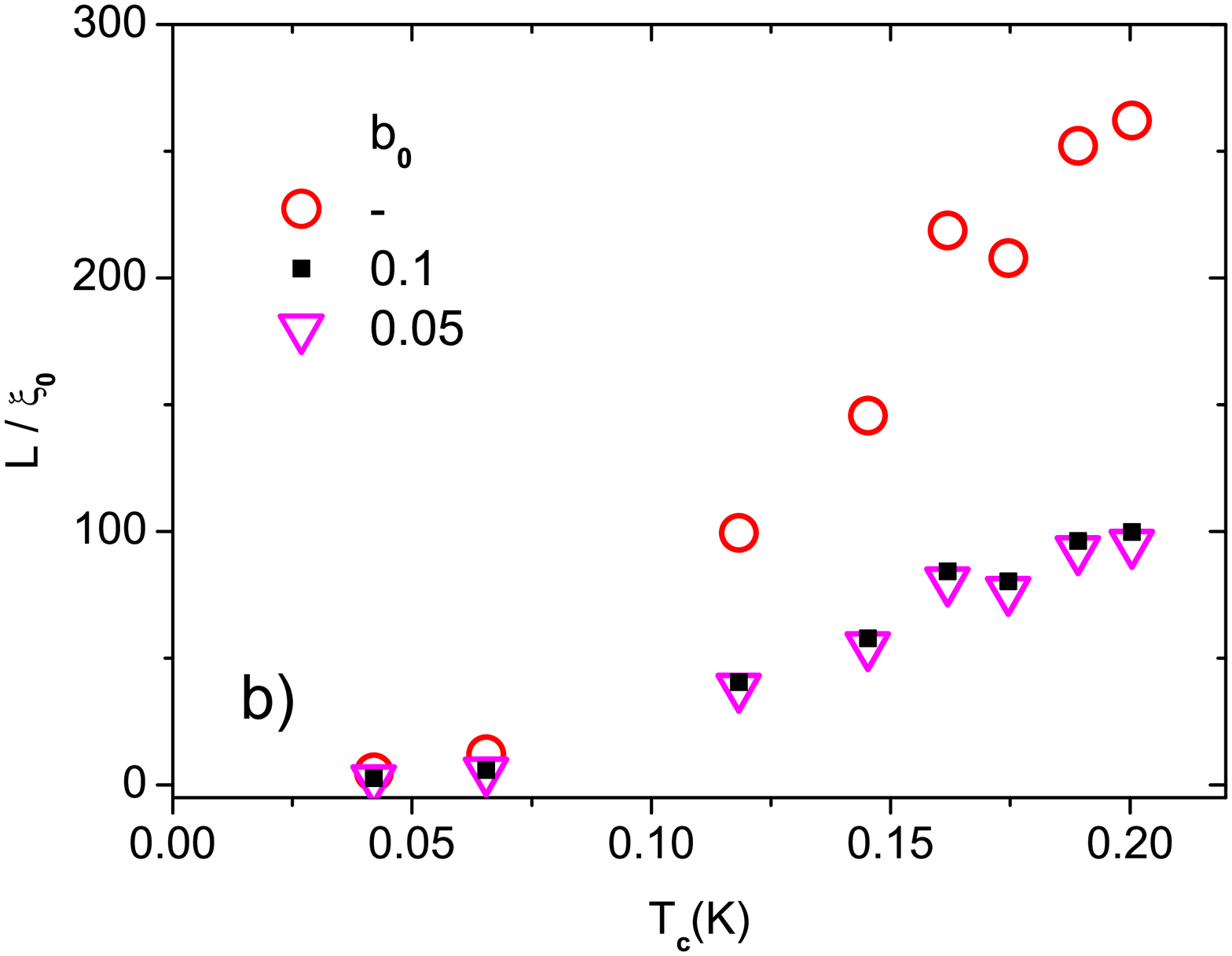} \vspace{-0.5cm}
\caption{(color online) (a) $R\left( V_{g},T\right)
/R_{0}\left( V_{g}\right) $ vs $T_{c}$ and gate voltage $V_{g}$ of the LaAlO%
$_{3}$/SrTiO$_{3}$ interface derived from the data shown in Fig. (\protect\ref{fig8}); (b)
Estimates for the ratio $L/\xi _{0}$ between correlation length and vortex
core radius without the multiplicative logarithmic correction term ($%
\bigcirc $) and with this correction for different $b_{0}$ values entering
Eq. \protect\ref{eq13}.}
\label{fig9}
\end{figure}

To unravel the consistency of the rounded transitions with a finite size
effect, we invoke Eq. (\ref{eq13}) to estimate the ratio $L_{\min }/\xi _{0}$%
. Fig. \ref{fig9}a shows the $T_{c}$ and $d$ dependence of $R\left(
V_{g},T_{c}\right) /R_{0}\left( V_{g}\right) $ derived from Fig. \ref{fig8}a. The
resulting $T_{c}$ dependence of $L/\xi _{0-}$ is shown in Fig. \ref{fig8}b for $%
b_{0}=0.05$ and $0.1$ in comparison with the absence of the multiplicative
logarithmic correction. Note that $b_{0}=0.05$ is comparable to $%
b_{0}\approx 0.07$, derived from large-scale numerical simulations.\cite%
{andersson} In analogy to the Bi-films, important features include the
substantial decline $L/\xi _{0}$ with decreasing $T_{c}$, and the comparably
low values of $L/\xi $, namely $L/\xi _{0}<100$ compared to the lower bound $%
L/\xi _{0}\gtrsim 10^{5}$ emerging from the $^{4}$He data shown in Fig. \ref{fig2}.
According to this and in analogy to the Bi-films the run away from
BKT behavior as observed in Fig. \ref{fig9} is attributable to a limiting length $L$
where the ratio $L/\xi _{0}$ decreases with reduced $T_{c}$. Nevertheless,
there is a temperature range where consistency with BKT behavior is
observed, but in a strict sense a normal state to superconductor BKT
transition is suppressed.

\begin{figure}[tbh]
%\centering
\includegraphics[width=1.0\linewidth]{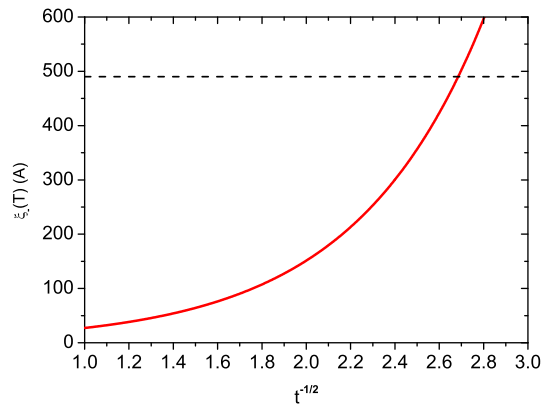} \vspace{-0.5cm}
\caption{(color online)Correlation length $\xi _{+}\left( T\right) =\xi _{0}\exp
\left( 2\pi /\left( bt^{1/2}\right) \right) $ vs $t^{-1/2}$ of the LaAlO$%
_{3}$/SrTiO$_{3}$ interface with $T_{c}\simeq 0.21$ K for $\xi _{0}=4.9$ {\AA}
and $2\pi /b=b_{R}/2=1.72$. The dashed line marks $L=208$ {\AA}. The crossing
point at $t^{-1/2}\simeq 2.69$ corresponds to $T_{c}/T\simeq 0.88$.}
\label{fig10}
\end{figure}

An independent confirmation of the finite size scenario demands the
magnitude of $L$, allowing to determine $\xi _{0}$ and with that the
temperature dependence of the correlation length $\xi _{+}\left( T\right) $,
as well as $\xi _{+}\left( T^{\ast }\right) $ $=L$, where the run away from
BKT behavior should occur. Given the previous estimate derived from the
magnetic field induced finite size effect\cite{tstool}

\begin{equation}
L\simeq 490\text{ {\AA},}  \label{eq19}
\end{equation}%
for a LaAlO$_{3}$/SrTiO$_{3}$ interface with $T_{c}\simeq 0.21$ we obtain
with $L/\xi _{0}\simeq 100$, taken from Fig. \ref{fig9}b, for the vortex core radius
the value%
\begin{equation}
\xi _{0}\simeq 4.9\text{ {\AA}.}  \label{eq20}
\end{equation}%
The resulting temperature dependence of the correlation length is shown in
Fig. \ref{fig10} in terms of $\xi _{+}\left( T\right) $ vs $t^{-1/2}$. As
the correlation length cannot grow beyond $L$ the run away from BKT behavior
should occur around the crossing point between $\xi _{+}\left( T\right) $
and $L$ at $t^{-1/2}\simeq 2.69$ corresponding to $T_{c}/T\simeq 0.88$. A
glance at Fig. \ref{fig8}a reveals that around this value the data of the LaAlO$_{3}$%
/SrTiO$_{3}$ interface at gate voltage $V_{g}=80$ V ($T_{c}\simeq 0.2$ K)
run away from the BKT behavior. This agreement reveals that magnetic field
and zero field finite size scaling yield consistent results. On this ground
is the smeared BKT transition in both, the Bi-films and the LaAlO$_{3}$/SrTiO%
$_{3}$ interface, attributable to a finite size effect stemming from a
limiting length $L$. In the samples with highest $T_{c}$ its dimension is $%
L\simeq 208$ {\AA} in the Bi-films and $L\simeq 490$ {\AA} in the LaAlO$_{3}$/SrTiO$%
_{3}$ interface.

\begin{figure}[tbh]
%\centering
\includegraphics[width=1.0\linewidth]{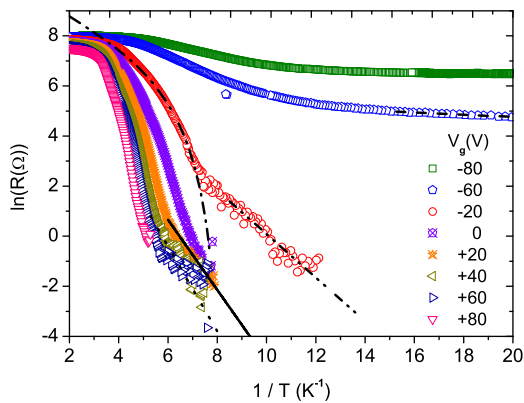} \vspace{-0.5cm}
\caption{(color online)  $\ln (R)$ vs $1/T$ of the LaAlO$_{3}$/SrTiO$_{3}$
interface for various gate voltages. The straight lines are Eq. (\protect\ref{eq21}%
): dashed line: $V_{g}=-60$ V with $r=5.64$ and $s\left( T\right) =0.044$ K;
dash-dot-dot line: $V_{g}=-20$ V with $r=8.78$ and $s\left( T\right) =0.87$
K; full line: $V_{g}=+20$ V with $r=9.06$ and $s\left( T\right) =1.4$ K;
dotted line: $V_{g}=+60$ V with $r=9.8$ and $s\left( T\right) =1.7$ K. The
beginnings of the lines mark the respective $1/T_{c}$. The dash-dot line
marks the BKT behavior (\protect\ref{eq6}) at $V_{g}=-20$ V with $R_{0}=44$ k$\Omega
$, $b_{R}=3.43$ and $T_{c}=0.119$ K.}
\label{fig11}
\end{figure}

Next we turn to the finite size behavior below the extrapolated transition
temperature. Here the limiting length $L$ prevents the thermal length $\xi
_{-}\left( \left\vert t\right\vert \right) $ to diverge. But compared to $%
\xi _{+}\left( \left\vert t\right\vert \right) $ the thermal length is much
smaller for the same $\left\vert t\right\vert $ (Eq. (\ref{eq12a})). For
this reason $L\gtrsim \xi _{-}\left( T\right) $ is expected to hold already
slightly below $T_{c}$. In this regime the sheet resistance is controlled by
the free vortex density where Eq. (\ref{eq11}) rewritten in the form%
\begin{equation}
\ln \left( R\left( T\right) \right) =r-\frac{s\left( T\right) }{T},s\left(
T\right) =\frac{\pi J\left( T\right) }{k_{B}}\ln \frac{L}{\xi _{0}}
\label{eq21}
\end{equation}%
applies. Accordingly, the coefficient $s\left( T\right) $ controls
deviations from the $1/T$ temperature dependence. At zero temperature the
superfluid stiffness given by Eq. (\ref{eq7a}) is fixed by the magnetic
penetration depth in terms of $J\left( T=0\right) \propto d/\lambda
^{2}\left( T=0\right) $, expected to vanish as $J\left( T=0\right) \propto
d/\lambda ^{2}\left( T=0\right) \propto T_{c}$.\cite{tswbol} On the other
hand, approaching $T_{c}$ from below, the superfluid stiffness tends
according to Eq. (\ref{eq10}) to $J\left( T_{c}^{-}\right) =2k_{B}T_{c}/\pi $%
. In addition in both, the Bi-films (Fig. \ref{fig4}b) and the LaAlO$_{3}$/SrTiO$_{3}$
interface (Fig. \ref{fig9}b)), ln$\left( L/\xi _{0}\right) $ decreases with reduced $%
T_{c}$. As a consequence the magnitude of $s\left( T\right) $ is expected to
decrease with reduced $T_{c}$. In Fig. \ref{fig11}, showing $\ln (R)$ vs $1/T$ of
the LaAlO$_{3}$/SrTiO$_{3}$ interface for various gate voltages, we observe
that this supposition is well confirmed. On the other hand, in the
temperature regime of interest the data exhibit jitter masking the
characteristic temperature dependence of the superfluid stiffness in $%
s\left( T\right) $.\cite{nelson} Indeed, the straight lines, corresponding
to the nearly temperature independent $s\left( T\right) \approx 2T_{c}\ln
\left( L/\xi _{0}\right) $, describes the data quite well. To evidence the
smeared BKT transition we included in Fig. \ref{fig11} the characteristic BKT
temperature dependence (\ref{eq6}) in terms of the dash-dot-dot line.
Additional confirmation of this finite size scenario below $T_{c}$ stems
from the observation of an ohmic regime at small currents\cite{reyren}
because it uncovers according to Eq. (\ref{eq1}) the presence of free
vortices. The important implication then is: although BKT behavior is
observable in an intermediate temperature regime above the extrapolated $%
T_{c}$, in a strict sense a BKT transition does not occur. It is smeared out
and the sheet resistance vanishes at zero temperature only because Eq. (\ref%
{eq21}) reduces in the zero temperature limit to%
\begin{eqnarray}
R\left( T\right)  &=&r\exp -\left( \frac{\pi J\left( T=0\right) }{k_{B}T}\ln
\frac{L_{\lim }}{\xi _{0}}\right)   \nonumber \\
&=&r\left( \frac{\xi _{0}}{L_{\min }}\right) ^{\frac{\pi J\left( T=0\right)
}{k_{B}T}}.  \label{eq22}
\end{eqnarray}

\begin{figure}[tbh]
%\centering
\includegraphics[width=1.0\linewidth]{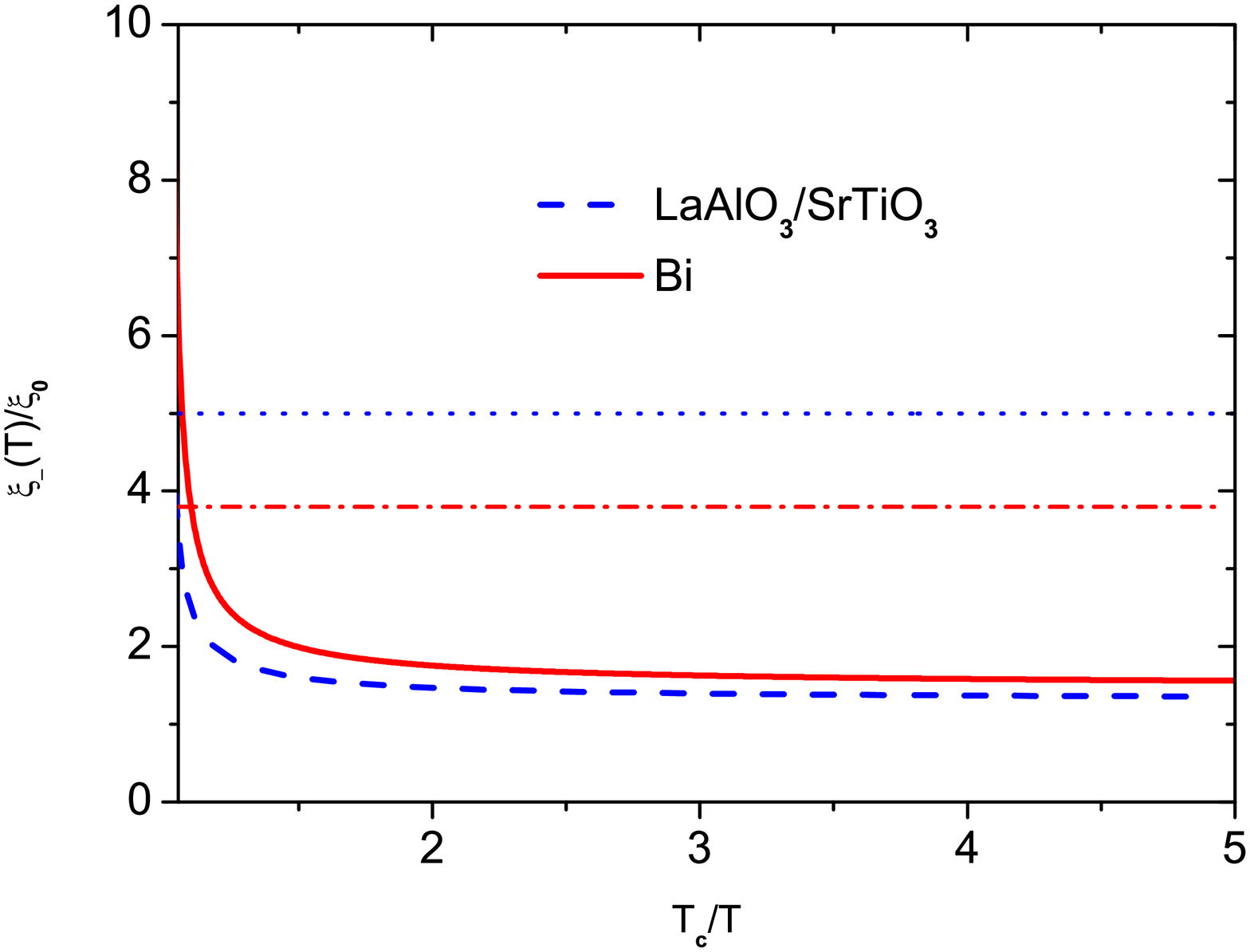} \vspace{-0.5cm}
\caption{(color online)  $\xi
_{_{-}}\left( T\right) /\xi _{0}=$exp$\left( 1/\left( b\left(
1-T/T_{c}\right) ^{1/2}\right) \right) $ vs $T/T_{c}$ for the Bi-films with
$1/b=b_{R}/4\pi \simeq 0.398$ and the LaAlO$_{3}$/SrTiO$_{3}$ interface with
$1/b=b_{R}/4\pi \simeq 0.273$. The dash dot and dotted lines mark the
minimum value of $L/\xi _{0}$. $L/\xi _{0}$ $\simeq 3.8$ for the Bi-films (Fig. \protect\ref{fig4}b) and $L/\xi _{0}$ $\simeq 5$ for the LaAlO$_{3}$/SrTiO$_{3}$
interface(Fig. \protect\ref{fig9}b).}
\label{fig12}
\end{figure}

Contrariwise, the sheet resistance of the Bi-films shown in Fig. \ref{fig3} does not
exhibit a significant temperature dependence below $T\approx T_{c}/2$ down
to $T\approx T_{c}/10$. To disentangle the scaling regimes below $T_{c}$
more quantitatively, we note that the plot $R\left( T\right) /R_{0}$ vs $%
T_{c}/T$ should exhibit a crossover from a temperature dependent to a
temperature independent regime at $T^{\ast }$ where the diverging length
$\xi _{_{-}}\left( T\right) $ equals the limiting length $L_{\min }$.
According to Eqs. (\ref{eq11}) and (\ref{eq12}) $T^{\ast }$ follows from
\begin{equation}
\frac{L}{\xi _{0}}=\frac{\xi _{_{-}}\left( T^{\ast }\right) }{\xi _{0}}=\exp
\left( \frac{1}{b\left( 1-T^{\ast }/T_{c}\right) ^{1/2}}\right) .
\label{eq23}
\end{equation}%
To estimate $T^{\ast }$ we show in Fig. \ref{fig12} the temperature dependence of $\xi _{_{-}}\left( T\right) $
in terms of $\xi _{_{-}}\left( T\right) /\xi
_{0}$ vs $T/T_{c}$ for the Bi-films and the LaAlO$_{3}$/SrTiO$_{3}$
interface. Noting that the minimum value of $L/\xi _{0}$ in the Bi-films is
around $3.8$ (Fig. \ref{fig4}b) and in the LaAlO$_{3}$/SrTiO$_{3}$ interface around
around $5$ (Fig. \ref{fig9}b) it becomes clear that in both systems $T^{\ast }$ is
close and slightly below $T_{c}$. As a result, the temperature regime where
$\ \xi _{-}\left( T\right) >L_{\lim }$ holds is restricted to temperatures
very close to $T_{c}$ only, while the regime where $\xi _{-}\left( T\right)
<L$ applies sets in slightly below $T_{c}$. It is the regime where the sheet
resistance adopts the characteristic temperature dependence given by Eq. (\ref{eq21}).
A glance at Fig. \ref{fig11}, showing $\ln (R)$ vs $1/T$ of the LaAlO$_{3}$/SrTiO$_{3}$ interface,
uncovers agreement with this temperature
dependence, while the sheet resistance of the Bi-films shown in Fig. \ref{fig3} does
not exhibit a significant temperature dependence below $T\approx T_{c}/2$.
Taking the saturation of the sheet resistance in the BI-films for granted it
implies the breakdown of the BKT behavior below $T_{c}$, while it applies
above $T_{c}$. The breakdown may then be a clue that below $T_{c}$ a process
is present which destroys BKT behavior. On the other hand we have seen that
the LaAlO$_{3}$/SrTiO$_{3}$ interface data is at and below $T_{c}$
remarkably consistent with the predicted finite size BKT behavior. However,
the absence of BKT behavior below $T_{c}$ is inconsistent with measurements
of the superfluid stiffness,\cite{misra,turneaure,bert} uncovering a smeared
Nelson-Kosterlitz\cite{nelson} jump near $T_{c}$ and the presence of
superfluidity down to the lowest attained temperatures. Given the odd
behavior of the Bi-films it should be kept in mind that a failure to cool
the electrons in the low temperature limit also implies a flattening of the
sheet resistance.\cite{parendo}

\begin{figure}[tbh]
%\centering
\includegraphics[width=1.0\linewidth]{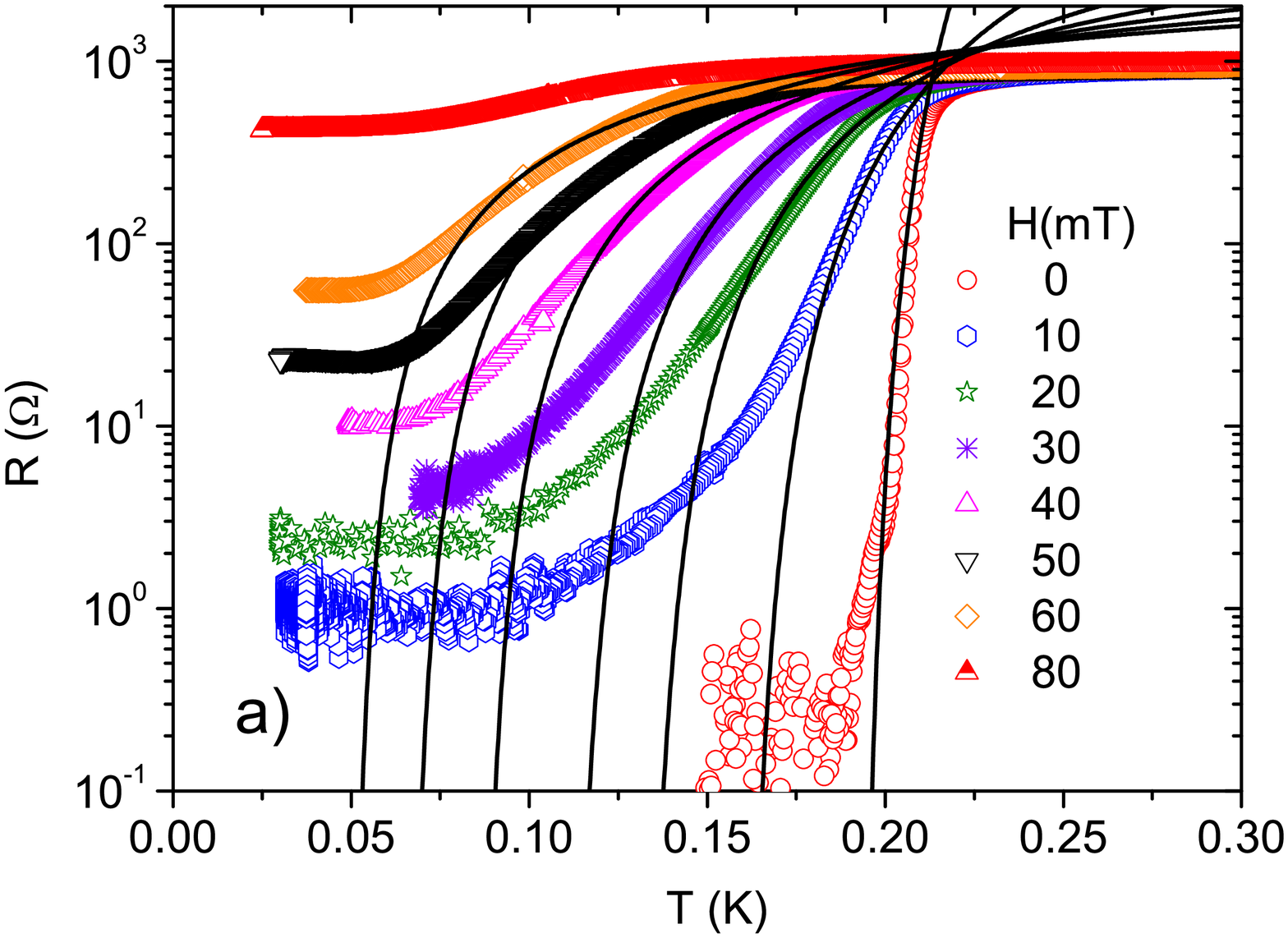} \vspace{-0.5cm}
\includegraphics[width=1.0\linewidth]{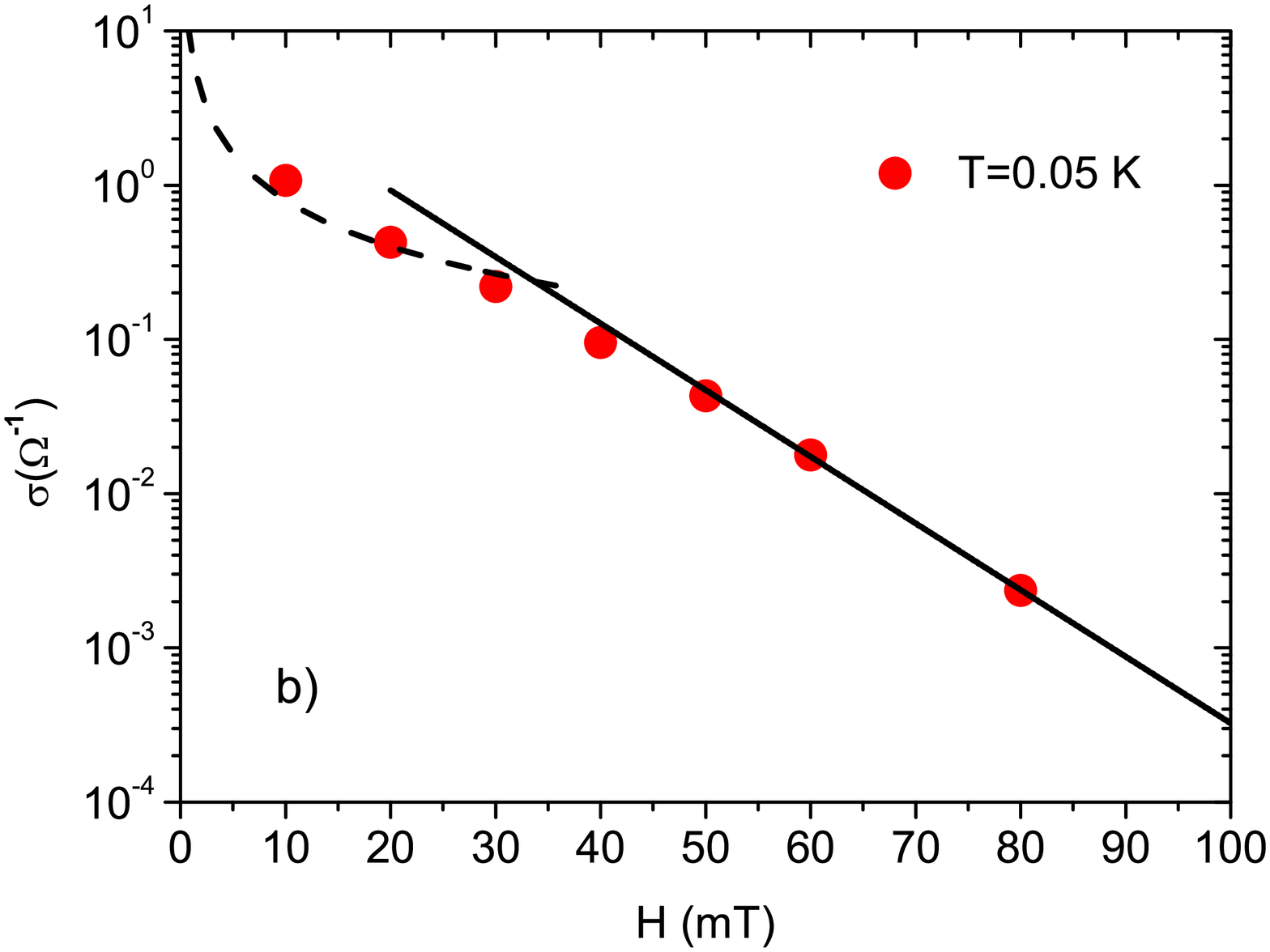} \vspace{-0.5cm}
\caption{(color online) (a) Temperature dependence of the sheet resistance of a LaAlO$_{3}$%
/SrTiO$_{3}$ interface with $T_{c}\simeq 0.19$ K at various magnetic fields
applied perpendicular to the interface taken from Reyren \textit{et al}.\protect\cite%
{reyren2} The solid lines are fits to the BKT form (\ref{eq6}) of the sheet
resistance with $b_{R}=3.43$ yielding for $T_{c}$ and $R_{0}$ the estimates
shown in Fig. \protect\ref{fig14} ; (b) Sheet conductivity vs $H$ at $%
T=0.05$ K. The solid line is the empirical form (\ref{eq18a}) with $%
\overline{\sigma }_{0}=6.79$ $\Omega ^{-1}$ and $\overline{a}=0.099$ mT$^{-1}
$. The dashed line is Eq. (\ref{eq15}) in the from $\sigma =\widetilde{%
\sigma }/H$ where $\widetilde{\sigma }=8$ ($\Omega ^{-1}$mT).}
\label{fig13}
\end{figure}

Finally, to explore the implications of a magnetic field induced finite size
effect below $T_{c}$ we depicted in Fig. \ref{fig13}a the temperature
dependence of the sheet resistance of a LaAlO$_{3}$/SrTiO$_{3}$ interface
with $T_{c}\simeq 0.19$ K at various magnetic fields. Although the data
exhibit jitter in the low field limit the predicted saturation of the sheet
resistance in the $T\rightarrow 0$ limit (Eq. (\ref{eq15})) is well
established. On the other hand, considering the isotherm shown in Fig. \ref%
{fig13}b, the consistency with the finite size behavior (\ref{eq15}) is
restricted to low temperatures and low fields. Above $H=30$ mT a crossover to
the empirical form (\ref{eq18a}) can be surmised as the crossing point of
the isotherms around $H_{c}=110$ mT is approached. This crossing point is
the direct consequence of the fact that in the covered $T$ range $R$
decreases with decreasing $T$ for $H<H_{c}$, increases with decreasing $T$
for $H>H_{c}$, and is $T$ independent at $H_{c}$. Noting that the scaling
form (\ref{eq15}) presumes that density fluctuations are small,\cite%
{andersson} which is true for large limiting lengths $L_{H}=\left( \Phi
_{0}/aH\right) ^{1/2}$, but not for small, it becomes clear that the
applicability of this approach is limited to the low field limit. Another
essential feature emerging from Fig. \ref{fig13}a is the shift of the sheet
resistance curves to lower temperatures with increasing magnetic field. This
behavior uncovers the pair breaking effect of the magnetic field leading in
a mean-field treatment to a reduction of $T_{c0}$ according to $T_{c0}\left(
H=0\right) -T_{c0}\left( H\right) \propto H$.\cite{abrikosov,aoi,shah} Adopting
the finite size point of view this behavior relies on the fact that an
applied magnetic field sets an additional limiting length $L_{H}=\left( \Phi
_{0}/aH\right) ^{1/2}$, giving rise to a smeared BKT transition at a
fictitious BKT transition temperature $T_{c}\left( H\right) $ below $%
T_{c}\left( H=0\right) $. Contrariwise, in the standard finite size effect
one attains $T_{c}$ in the $L\rightarrow \infty $ limit only. To quantify
this option we performed fits to the characteristic BKT form (\ref{eq6}) of
the sheet resistance. A glance at Fig. \ref{fig13}a reveals, in analogy to the
zero field case (Fig. \ref{fig8}a), agreement in an intermediate temperature
range below $T_{c}\left( H\right) $.

\begin{figure}[tbh]
%\centering
\includegraphics[width=1.0\linewidth]{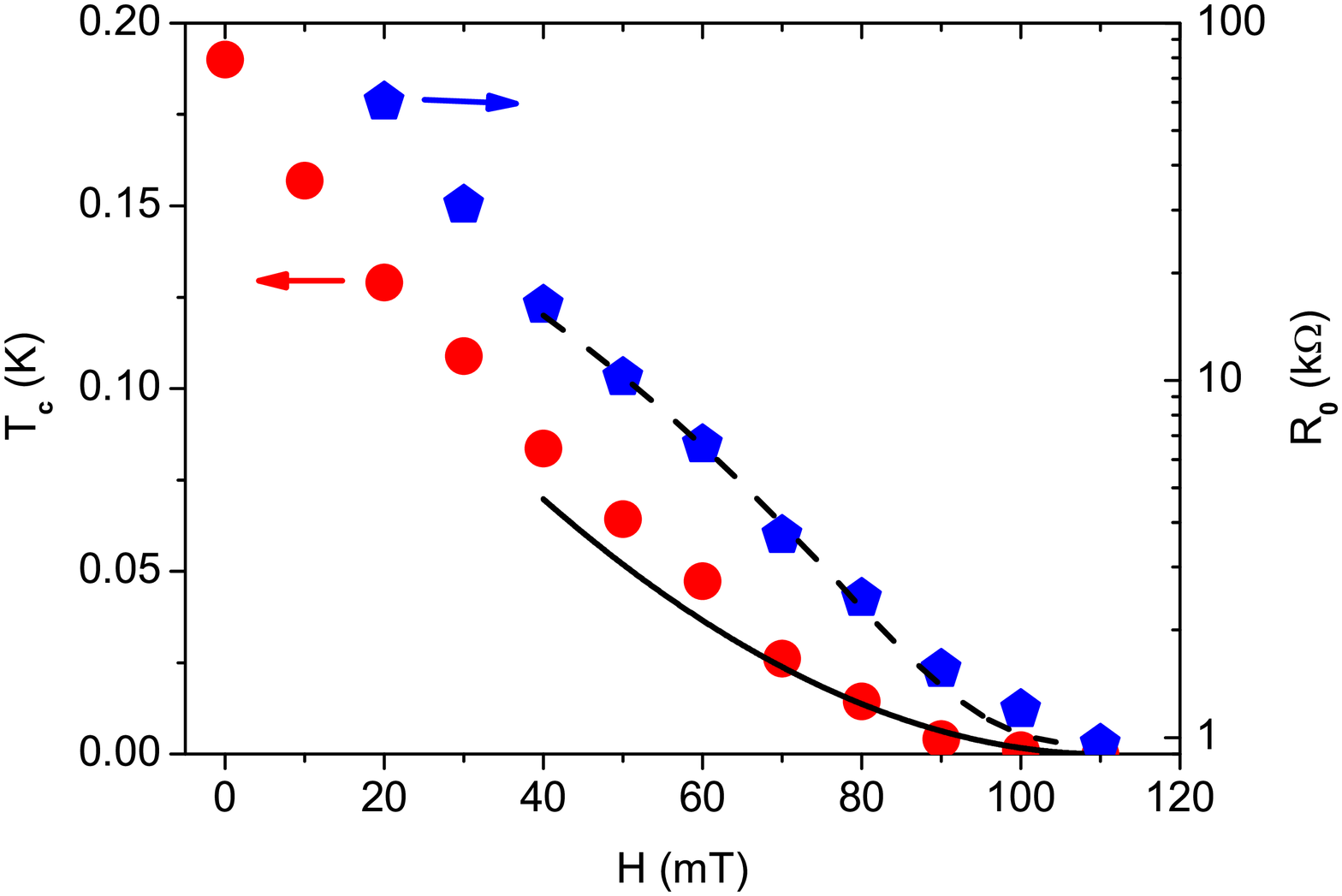} \vspace{-0.5cm}
\caption{(color online)    Estimates for $T_{c}$ and $R_{0}$ resulting from the fits to the
BKT form (\protect\ref{eq6}) of the sheet resistance included in Fig. \protect\ref{fig13}a.
The solid line is $T_{c}$=$T_{0}(H_{c}-H)^{z\overline{\nu }}$ with $%
T_{0}=3\cdot 10^{-5}$ $\left( \text{KmT}\right) ^{1/z\overline{\nu }}$, $%
H_{c}=110$ mT, $z\overline{\nu }=1.92\pm 0.1$ and the dashed line is $%
R_{0}=R_{0c}+\overline{R}(H_{c}-H)^{2\overline{\nu }}$ with $R_{0c}=0.96$ k$%
\Omega $ , $\overline{R}=0.106$ $\Omega $mT$^{1/2\overline{\nu }}$, and $2%
\overline{\nu }=2.78$. These lines indicate the approach to the extrapolated
quantum critical point.}
\label{fig14}
\end{figure}

Given the consistency with the BKT expression (\ref{eq6}) and Fig. (\ref{fig13}a)
estimates for the fictitious lines $T_{c}\left( H\right) $
and $R_{0}\left( H\right) $ are readily obtained and shown in Fig. \ref{fig14}.
$T_{c}\left( H\right) $ extrapolates to zero around $H_{c}=110$ mT where the
isotherms cross. This behavior suggests a magnetic field induced quantum
phase transition where superconducting behavior is lost at zero temperature
and the amplitude $R_{0}$ approaches the critical value $R_{0}\left(
H_{c}\right) \simeq 1$ k$\Omega $ which is close to the normal state sheet
resistance at $T=0.5$ K. We note that $T_{c}\left( H\right) $ has properties
compatible with a quantum critical point, where $T_{c}$=$T_{0}(H_{c}-H)^{z\overline{\nu }}$ applies.\cite{sondhi}
$z$ is the dynamic and $\overline{\nu }$ the critical exponent of the zero temperature correlation length. The
power law fit included in Fig. \ref{fig14} yields $z\overline{\nu }=1.92\pm 0.1$.
It is interesting to note that this value is comparable with transport
studies including MoGe,\cite{mason} Nb$_{0.15}$Si$_{0.85}$,\cite{aubin}, InO$_{x}$, \cite{hebard2},
and LaAlO$_{3}$/SrTiO$_{3}$ interface\cite{biscaras}
samples, though these studies have limited their analysis to exclude
resistance data showing flattening in the zero temperature limit. In any
case, BKT behavior occurs in an intermediate temperature range only. The
extrapolated BKT line $T_{c}\left( H\right) $ is not attainable because the
magnetic field induced finite size effect (Eq. \ref{eq15})) generates, as
observed in Fig. \ref{fig13}a, the flattening out of the sheet resistance in the
$T\rightarrow 0$ limit. Nevertheless, the established survival of
BKT behavior in a magnetic field applied perpendicular to the film also
implies a smeared sudden drop in the superfluid stiffness at $T_{c}\left(
H\right) $, where the superfluid stiffness adopts the universal value given
by the Nelson-Kosterlitz relation (\ref{eq10}). Recently, this behavior has
been observed in MoGe and InO$_{x}$ thin films by means of low frequency
measurements of the ac conductivity.\cite{misra} Although the low frequency $%
f=20$ kHz implies an additional limiting length, namely $L_{f}\propto
f^{-1/2}$, the magnetic field dependence of the blurred Nelson-Kosterlitz
jump has been clearly detected and the power law fits to $T_{c}\left(
H\right) $ yielded for $z\overline{\nu }$ the estimates $1.25\pm 0.25$ for
MoGe and $1.3\pm 0.4$ for InO$_{x}$.

\begin{figure}[tbh]
%\centering
\includegraphics[width=1.0\linewidth]{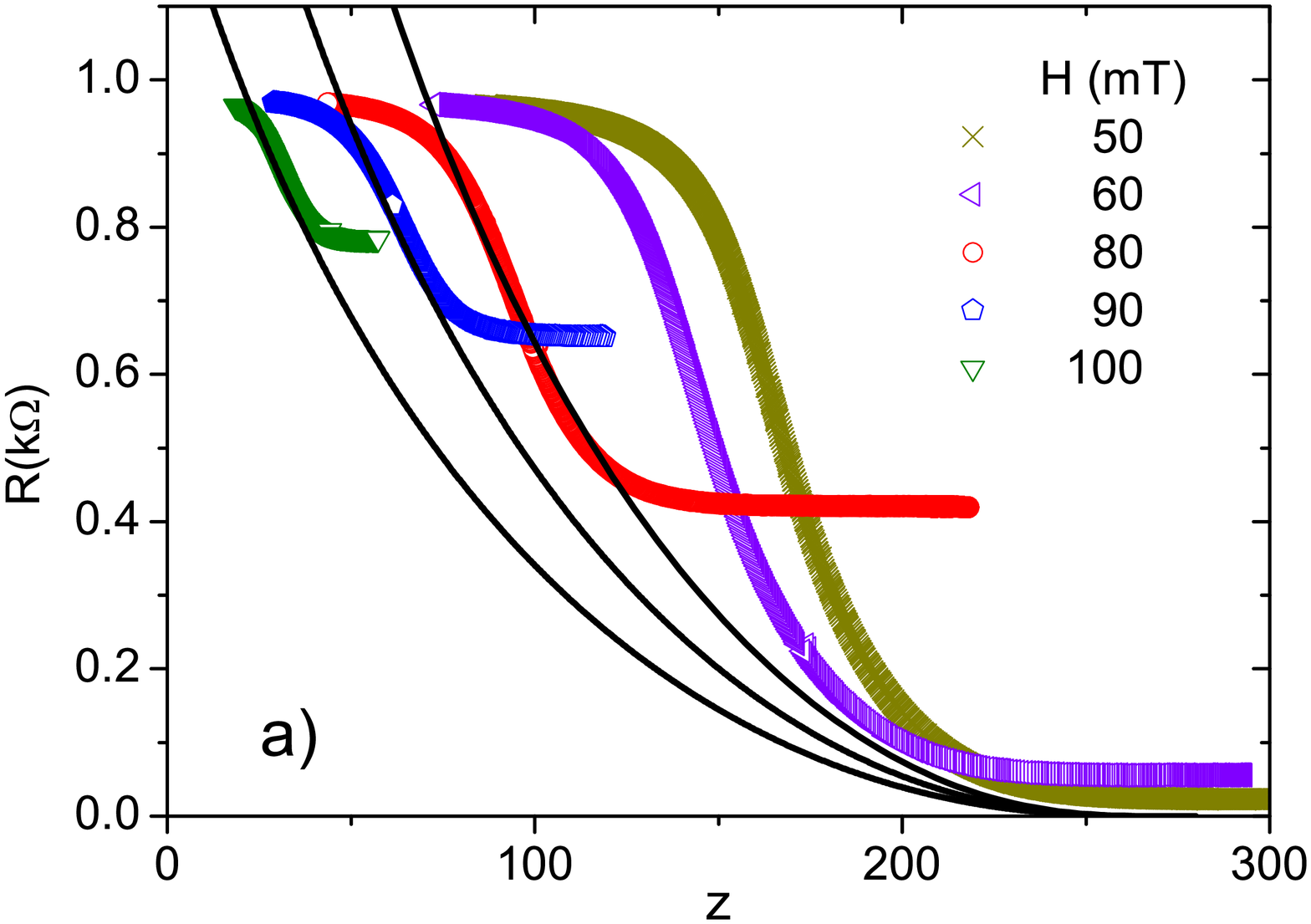} \vspace{-0.5cm}
\includegraphics[width=1.0\linewidth]{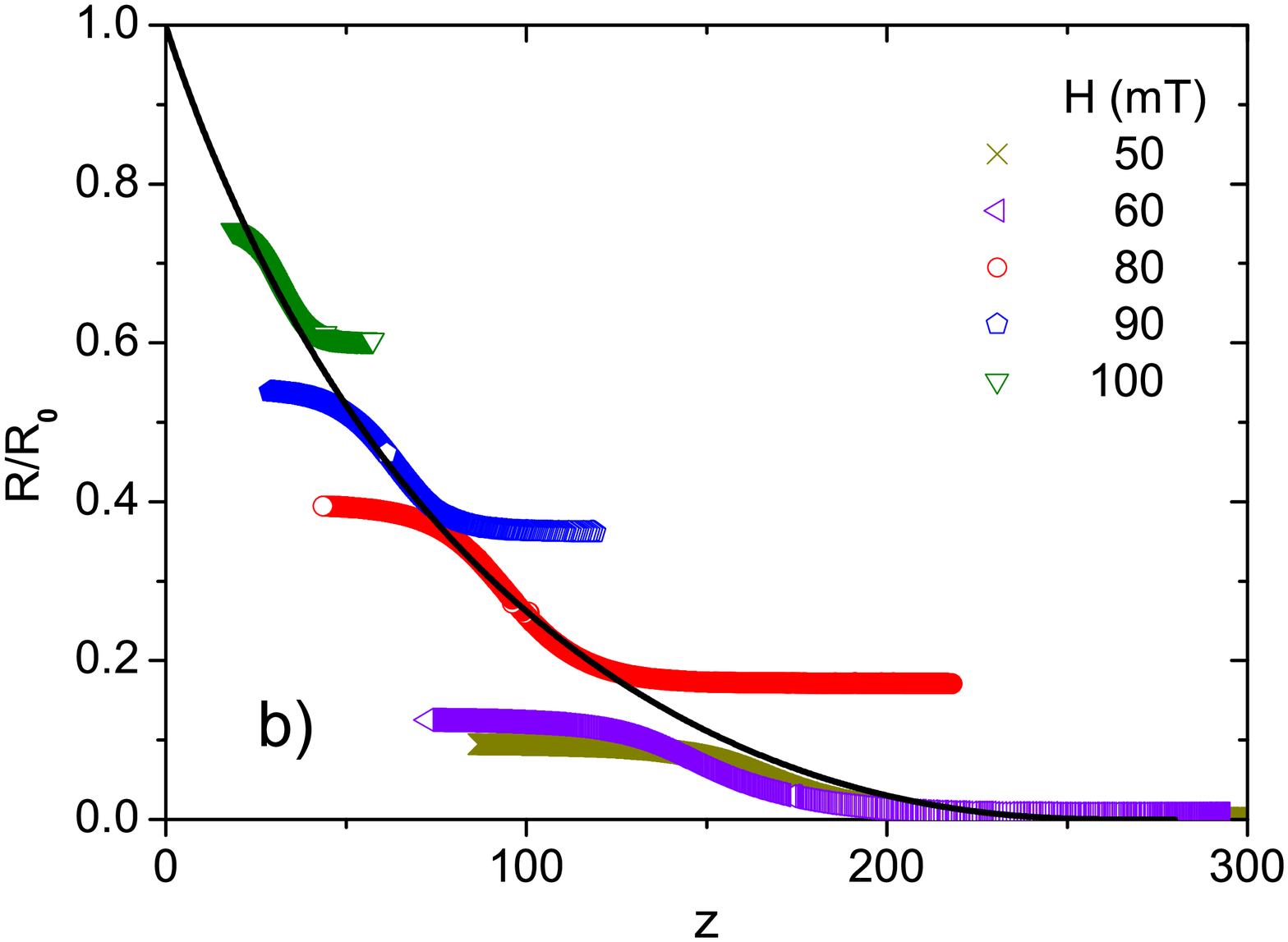} \vspace{-0.5cm}
\caption{(color online)  (a) Scaling plot $R\left( H,T\right) $ vs $%
z=\left( H_{c}-H\right) /T^{1/z\overline{\nu }}$ with $H_{c}=110$ mT, $z%
\overline{\nu }=1.92$, and $b_{R}$ $=3.43$. The solid lines mark the
respective BKT scaling form (\protect\ref{eq27}) with $R_{0}\left( H\right) $ taken
from Fig. \protect\ref{fig14} and $T_{0}=2\cdot 10^{-5}$ $\left( \text{KmT}\right)
^{1/z\overline{\nu }}$. (b) Scaling plot $R\left( H,T\right) /R_{0}\left(
H\right) $ vs $z=\left( H_{c}-H\right) /T^{1/z\overline{\nu }}$.
The solid line is the BKT scaling form (\protect\ref{eq27}).}
\label{fig15}
\end{figure}

Lastly we consider the limitations of the quantum scaling form\cite{sondhi}
\begin{equation}
R\left( H,T\right) =R_{c}G\left( x\right) ,x=\frac{H_{c}-H}{T^{1/z\overline{%
\nu }}},  \label{eq24}
\end{equation}%
applicable close to the quantum critical point. $G\left( x\right) $ is a
scaling function of its argument and $G\left( 0\right) =1$. It is
essentially a finite size scaling function. Indeed at finite temperatures is
the divergence of the zero temperature correlation length $\xi \left(
T=0\right) \propto \left( H_{c}-H\right) ^{-\overline{\nu }}$ cutoff by the
thermal length $L_{T}\propto T^{-1/z}$ so that $x\propto \left( L_{T}/\xi
\left( T=0\right) \right) ^{1/\overline{\nu }}\propto \left( H_{c}-H\right)
/T^{1/z\overline{\nu }}T$. The data for $R\left( H,T\right) $ plotted
vs $x=\left( H_{c}-H\right) /T^{1/z\overline{\nu }}$ should then
collapse on a single curve. On the other hand BKT behavior uncovered in Fig. \ref{fig13}a
implies the scaling form (\ref{eq6}) rewritten in the form%
\begin{equation}
R\left( H,T\right) =R_{0}\left( H\right) \exp \left( -b_{R}/\left(
T/T_{c}\left( H\right) -1\right) ^{1/2}\right) ,  \label{eq25}
\end{equation}%
where $T_{c}\left( H\right) $=$T_{0}(H_{c}-H)^{z\overline{\nu }}$ is the
transition line shown in Fig. \ref{fig14}. Noting that

\begin{equation}
\frac{T}{T_{c}\left( H\right) }=\frac{1}{T_{0}x^{z\overline{\nu }}},
\label{eq26}
\end{equation}
BKT behavior leads with Eqs. (\ref{eq24}) and (\ref{eq25}) to the explicit
scaling form%
\begin{equation}
R\left( H,T\right) =R_{0}\left( H\right) \exp \left( -b_{R}/\left( \left(
T_{0}x^{z\overline{\nu }}\right) ^{-1}-1\right) ^{1/2}\right) ,  \label{eq27}
\end{equation}%
valid for any $T/T_{c}\left( H\right) =\left( T_{0}x^{z\overline{\nu }%
}\right) ^{-1}>1$ where the universal critical behavior is entirely
classical. The scaling plot $R\left( H,T\right) $ vs $z=\left(
H_{c}-H\right) /T^{1/z\overline{\nu }}$ obtained from the LaAlO$_{3}$/SrTiO$_{3}$
interface sheet resistance data shown in Fig. \ref{fig13}a is depicted
in Fig. \ref{fig15}a. For comparison we included the BKT scaling form (\ref%
{eq27}). Apparently, the data do not collapse on a single curve because the
amplitude  $R_{0}$ exhibits a pronounced field dependence (see Fig. \ref%
{fig14}) and the sheet resistance flattens out for large and small values of
the scaling argument $z$. For fixed $H_{c}-H$ this reflects the observed
flattening out of the sheet resistance in the low and high temperature
limits (Fig. \ref{fig13}a). A glance at Fig. \ref{fig15}b reveals that an
improved data collapse is achieved by taking the field dependence of the
amplitude $R_{0}$ into account. Clearly, the flattening out for small and
large $z$ values remains. Noting that for fixed $H_{c}-H$ small $z$ values
are attainable at rather high temperatures only, the respective
saturation reflects the fact that in this temperature regime BKT
fluctuations no longer dominate. On the other hand large scaling arguments
require the incidence of the zero temperature limit where the magnetic field
induced finite size effect leads to a flattening out in the temperature
dependence and with that in the $z$ dependence of the sheet resistance in
the $z\rightarrow \infty $ limit. Furthermore, the field dependence of the
amplitude $R_{0}$ also implies that the quantum scaling form holds in a
unattainable regime close to quantum criticality only.

\section{Summary and discussion}

We analyzed sheet resistance data of thin Bi-films \cite{goldbi} and the
LaAlO$_{3}$/SrTiO$_{3}$ interface \cite{caviglia,reyren2} near the onset of
superconductivity to explore the compatibility with BKT behavior. On the
Bi-films the onset temperature has been tuned by the film thickness, while
on the LaAlO$_{3}$/SrTiO$_{3}$ interface the gate voltage and the magnetic
field, applied perpendicular to the interface, acted as tuning parameter.
Noting that BKT behavior involves the transition from a low-temperature
state in which only paired vortices exist to a high-temperature state in
which free vortices occur, we demonstrated that finite size induced free
vortices below $T_{c}$ prevent the occurrence of a BKT transition in a
strict sense. This does not mean, however, that the BKT vortex-unbinding
mechanism does not occur and is not observable. Indeed our finite size
analysis revealed that BKT behavior is present in an intermediate
temperature range above the extrapolated BKT transition temperature. This
temperature range depends on the magnitude of the limiting length $L$ while
the extrapolated transition temperature corresponds to the limit $%
L\rightarrow \infty $. Limiting lengths include he effective magnetic
penetration depth $\lambda _{2D}=\lambda ^{2}/d$ , the dimension $L_{h}$ of
the homogeneous domains in the sample, the magnetic length $L_{H}\propto
\left( \Phi _{0}/H\right) ^{1/2}$, and in the case of ac measurements $L_{f}\propto f^{-1/2}$.
$L=$min$\left[ \lambda _{2D},L_{h},L_{H},L_{f}\right]$
controls the density of free vortices $n_{_{F}}$ which determines the
sheet resistance ($R\propto n_{_{F}}$) as well as the correlation length
($\xi \propto n_{_{F}}^{-1/2}$) at and above $T_{c}$. In this temperature
range the limiting lengths prevent the correlation length to diverge.
Concentrating on the dc sheet resistance we analyzed the data using finite
size scaling formulas appropriate for the BKT transition.\cite{tstool,andersson}

The main results for zero magnetic fields include: Above $T_{c}$ we observed
in an intermediate temperature range consistency with the characteristic BKT
behavior and a thickness or gate voltage dependent BKT transition
temperature $T_{c}$ (Figs. \ref{fig1}a and \ref{fig8}a). However, in analogy to finite
systems, the measured sheet resistance does not vanish at $T_{c}$. In this
context it should be kept in mind that there is the Harris criterion,\cite{harris,aharony}
stating that short-range correlated and uncorrelated
disorder is irrelevant at the unperturbed critical point, provided that $\nu >2/D$,
where $D$ is the dimensionality of the system and $\nu$
the critical exponent of the finite-temperature correlation length. With $D=2$ and $\nu =\infty$,
appropriate for the BKT transition,\cite{kost}
this disorder should be irrelevant. Accordingly, the nonvanishing sheet
resistance at $T_{c}$ points to a
finite size induced smeared BKT transition. Invoking the finite size scaling
formula for the sheet resistance at $T_{c}$ we obtained estimates for the $T_{c}$
dependence of the ratio between the limiting length and the vortex
core radius, namely $L/\xi _{0}$ (Figs. \ref{fig3}b and \ref{fig9}b). Striking features
included the substantial decline of
$\left. L/\xi _{0}\right\vert _{\max}\approx 10^{2}$ with decreasing $T_{c}$ and in comparison with
$L/\xi_{0}\gtrsim 10^{5}$ in $^{4}$He the low value of $\left. L/\xi_{0}\right\vert _{\max }$.
This difference and the $T_{c}$ dependence of $L/\xi _{0}$ imply enhanced smearing of the BKT transition with reduced
$T_{c}$ as observed (Figs. \ref{fig1}a, \ref{fig3}, and \ref{fig8}a). To disentangle the $T_{c}$ dependence
of the limiting length $L$ and the vortex core radius $\xi _{0}$ we invoked
the magnetic field induced finite size effect allowing to estimate the
limiting length directly from magnetic field dependence of the sheet
conductivity at fixed temperature below $T_{c}$.\cite{tstool} Unfortunately,
in both the Bi-films and the LaAlO$_{3}$/SrTiO$_{3}$ interface, the
necessary data is available for the samples with highest $T_{c}$ only. For
the $23.42$ {\AA} thick Bi film we obtained $L\simeq 208$ {\AA}, $\xi _{0}\simeq 6.5$
{\AA} (Eqs. (\ref{eq17}) and (\ref{eq18})) and for the LaAlO$_{3}$/SrTiO$_{3}$
interface with $T_{c}\simeq 0.21$ K the estimates $L\simeq 490$ {\AA}, $\xi
_{0}\simeq 4.9$ {\AA} (Eqs. (\ref{eq19}) and (\ref{eq20})). These values for the
extent of the homogeneous domains are comparable with the dimension of the
superconducting patches emerging from scanning tunneling spectroscopy
investigations on TiNi\cite{sacepe} and InO$_{x}$\cite{sacepe2} films, as well as
with scanning SQUID measurements at the interface LaAlO$%
_{3}$/SrT iO$_{3}$.\cite{bert2} To
disentangle the $T_{c}$ dependence of $L$ and $\xi _{0}$ we used the
empirical relationship $\xi _{0}\propto T_{c}^{-1/z}$ with $z=2$,\cite{finotello,williams}
revealing that the extent of the homogenous domains
decreases substantially with reduced $T_{c}$ (Fig.\ref{fig7}). Accordingly the
enhanced smearing of the BKT transition with reduced $T_{c}$ was traced back
to the reduction of the limiting length $L$\ and the increase of the vortex
core radius $\xi _{0}$ with decreasing $T_{c}$.

In the low temperature limit and zero magnetic field we observed on the LaAlO%
$_{3}$/SrTiO$_{3}$ interface consistency with the characteristic finite size
scaling form (\ref{eq21}) while the Bi-films do not exhibit a significant
temperature dependence below $T\approx T_{c}/2$. Taking the saturation of
the sheet resistance in the BI-films for granted it implies the breakdown of
BKT finite size scaling below $T_{c}$, while it applies above $T_{c}$. The
breakdown may then be a clue that below $T_{c}$ a process is present which
destroys BKT behavior. On the other hand we have seen that the LaAlO$_{3}$%
/SrTiO$_{3}$ interface data is at and below $T_{c}$ remarkably consistent
with the predicted finite size BKT predictions. In addition, an absence of
BKT-behavior below $T_{c}$ is also incompatible with measurements of the
superfluid stiffness,\cite{turneaure,misra,bert} uncovering a smeared
Nelson-Kosterlitz\cite{nelson} jump near $T_{c}$ and the presence of
superfluidity down to the lowest attained temperatures.

Subsequently we explored the implications of the magnetic field induced finite
size effect. Considering the temperature dependence of the sheet resistance
at various magnetic fields, applied perpendicular to the interface of LaAlO$%
_{3}$/SrTiO$_{3}$, we observed in an intermediate temperature range
remarkable consistency with the characteristic BKT form (\ref{eq6})(Fig. \ref{fig13}a).
Fits yielded the fictitious transition line $T_{c}\left( H\right) $
extrapolating to zero at $H_{c}\simeq 110$ mT where a quantum phase
transition is expected to occur (Fig. \ref{fig14}). Indeed, $T_{c}\left( H\right) $
revealed properties compatible with a quantum critical point,
near which $T_{c}$=$T_{0}(H_{c}-H)^{z\overline{\nu }}$ applies.\cite{sondhi} $z$ is the
dynamic and $\overline{\nu }$ the critical exponent of the zero temperature
correlation length. A power law fit yielded $z\overline{\nu }=1.92\pm 0.1$.
However, this extrapolated line is not attainable because
the magnetic field induced finite size effect (Eq. (\ref{eq15}))
generates the observed flattening out of the sheet resistance in the $%
T\rightarrow 0$ limit (Fig. \ref{fig13}b). This feature has been observed in the
23.42 {\AA} thick Bi-film as well.\cite{goldbi} The survival of BKT behavior in
applied magnetic fields also implies a smeared sudden drop in the superfluid
stiffness at $T_{c}\left( H\right) $, where it adopts the universal value
given by the Nelson-Kosterlitz relation (\ref{eq10}). Recently, this
behavior has been observed in MoGe and InO$_{x}$ thin films by means of low
frequency measurements of the ac conductivity.\cite{misra}

A key question our analysis raises is whether the homogeneity of 2D
superconductors can be improved to reach the quality of $^{4}$He films.
Analyzing the sheet resistance data of Bi-films and the LaAlO$_{3}$/SrTiO$%
_{3}$ we have shown that the data are consistent with a finite size
effect attributable to the limited homogeneity of the samples. The limited
length of the homogenous domains impedes the occurrence of a BKT and quantum
phase transitions in the strict sense of a true continuous phase transition.
However, this strict interpretation of the definition of a continuous phase
transition does not imply that the BKT vortex-unbinding mechanism is not
observable and the reduction of the extrapolated $T_{c}$ does not reveal
properties compatible with a quantum critical point. Indeed, notwithstanding
the comparatively small dimension of the homogeneous domains, our finite
size analysis revealed reasonable compatibility with BKT and quantum
critical point behavior. However, the reduction of the limiting length with
decreasing $T_{c}$ is an essential drawback (Fig. \ref{fig7}).
Furthermore, considering the expected magnetic field tuned quantum phase
transition in the LaAlO$_{3}$/SrTiO$_{3}$ interface, it was shown that the
standard quantum scaling form (\ref{eq24}) of the sheet resistance applies
very close to the unattainable quantum critical point only (Fig. \ref{fig15}).
Indeed, combining the BKT expression for the sheet resistance with the
quantum scaling form of the extrapolated transition line $T_{c}\left(
H\right) $, we derived the explicit scaling relation (\ref{eq27}) uncovering
the limitations of the standard quantum scaling form. Its main drawback was
traced back to the neglect of the magnetic field dependence of the critical
amplitude $R_{0}$ which varies substantially by approaching the critical
value $R_{0c}$(Fig. \ref{fig14}).

Finally it should be noted that the finite size scaling approach adopted
here is compatible with the Harris criterion,\cite{harris,aharony} stating
that short-range correlated and uncorrelated disorder is irrelevant at the
BKT critical point, contrary to approaches where the smearing of the BKT
transition is attributed to a Gaussian-like distribution of the bare
superfluid-stiffness around a given mean value.\cite{benefatto}
The irrelevance of this disorder implies, that the universal BKT properties
still apply, while the nonuniversal parameters, including $T_{c}$, the
vortex core radius $\xi _{0}$ and the amplitude $R_{0}$, may change.
Contrariwise, the relevance of disorder at the extrapolated quantum phase
transition, separating the superconducting and metallic phase, depends on
the universality to which it belongs. The relevance of disorder is again
controlled by the Harris criterion:\cite{harris,aharony} if the
zero-temperature correlation length critical exponent fulfils the Harris
inequality $\overline{\nu }>2/D=1$ the disorder does not affect the quantum
critical behavior. Conversely, if $\overline{\nu }<2/D=1$ disorder is
relevant and affects the nonuniversal parameters $R_{0}$ and $T_{c}$ in the
BKT form (\ref{eq2}) of the sheet resistance and in particular the reduction
of $T_{c}$. In the magnetic field tuned case is the field
dependence of $R_{0}$ and $T_{c}$ attributable to Cooper pair breaking.
However, another important feature of the of LaAlO$_{3}$/SrTiO$_{3}$
interface is the large Rashba spin orbit interaction which originates from
the broken inversion symmetry. It has been shown that its magnitude
increases with reduced $T_{c}$,\cite{caviglia2} suggesting that pair breaking
occurs in zero magnetic field as well. Indeed,torque magnetometry
measurement revealed that the LaAlO$_{3}$/SrTiO$_{3}$ interface has a
magnetic moment, which points in the plane, and has an onset temperature
that is at least as high as 40 K and persists below the BKT transition
temperature.

\bigskip

\section{Acknowledgements}

The authors acknowledge stimulating and helpful discussions with K. A. M\"{u}ller and very useful comments from Stefano Gargilio.

\end{document}